\documentclass[fleqn,usenatbib]{mnras}

\usepackage{newtxtext,newtxmath}
\usepackage[T1]{fontenc}
\usepackage[normalem]{ulem}
\DeclareRobustCommand{\VAN}[3]{#2}
\let\VANthebibliography\thebibliography
\def\thebibliography{\DeclareRobustCommand{\VAN}[3]{##3}\VANthebibliography}
\usepackage{graphicx}
\usepackage{amsmath}
\usepackage{xcolor}
\usepackage{tikz}
\usetikzlibrary{shapes.geometric, arrows.meta, positioning, fit, backgrounds}
\usetikzlibrary{matrix, positioning, fit, backgrounds, arrows.meta}
\usepackage[percent]{overpic}
\usepackage{booktabs} 
\usepackage{multirow}
\usepackage{subcaption}
\usepackage{makecell}
\usepackage{arydshln}
\usepackage{placeins}
\usepackage{orcidlink}
\usepackage{bm}

\date{}
\makeatletter
\def\@printed{
  \qquad\qquad\qquad Compiled using MNRAS \LaTeX\ style file v\@version}
\makeatother

\title[OCCAMS: CMB with VAEs]{OCCAMS: A Multi-Task Learning-Based Investigation of the CMB Power Spectra}

\author[A. Mandal, R. Shah, M. W. Hossain, S. Sen, A. A. Sen, S. Pal]{
Ankit Mandal,$^{1}$\thanks{E-mail: mandalankit16@gmail.com \orcidlink{0009-0004-0001-0142}}
Rahul Shah,$^{1}$\thanks{E-mail: rahul.shah.13.97@gmail.com \orcidlink{0000-0001-7682-9219}}
Md. Wali Hossain,$^{2}$\thanks{E-mail: mhossain@jmi.ac.in \orcidlink{0000-0001-6969-8716}}
Somasri Sen,$^{2}$\thanks{E-mail: ssen@jmi.ac.in \orcidlink{0000-0001-8449-7242}}
Anjan A Sen,$^{3}$\thanks{E-mail: aasen@jmi.ac.in \orcidlink{0000-0001-9615-4909}}
Supratik Pal$^{1}$\thanks{E-mail: supratik@isical.ac.in \orcidlink{0000-0003-4136-329X}}
\\\\
$^{1}$Physics and Applied Mathematics Unit, Indian Statistical Institute, 203 B.T. Road, Kolkata 700 108, India\\
$^{2}$Department of Physics, Jamia Millia Islamia, New Delhi 110025, India\\
$^{3}$Centre for Theoretical Physics, Jamia Millia Islamia, New Delhi 110025, India
}

\begin{document}

\label{firstpage}
\pagerange{\pageref{firstpage}--\pageref{lastpage}}
\maketitle

\begin{abstract}
How many latent variables are needed to fit the cosmic microwave background (CMB) temperature and polarisation power spectra when a neural network is given no cosmological model, and to what extent do they map onto the standard cosmological parameters? We use multi-task learning with variational autoencoders trained on \texttt{CLASS}-generated spectra (TT, TE, EE) over $2\le\ell\le2500$. No cosmological model is built into the network, and the number of latent variables, $L$, is varied over $L=5$, $6$ and $7$. Their fits are compared to the Planck 2018 data using $\hat{\chi}^2$, AIC, BIC and Bayesian evidence. Their correspondence with cosmological parameters is measured with normalised mutual information post-training. The loss weights are set within a single training run by uncertainty weighting, with the weight of the disentanglement term either fixed (semi-automated, SA) or learned (completely automated, CA). The reconstruction loss is split over multipole ranges in three configurations that separate the roles of temperature and polarisation and of the low- and high-$\ell$ ranges. In four of the six combinations of configuration and strategy, the fit improves from $L=5$ to $L=6$ and not beyond, and $L=6$ is the best, or close to the best, in five of the six cases. So, six latent variables are needed, as many as the free parameters of the standard, model-dependent approach. SA gives clearer disentanglement, while CA needs no \textit{a priori} choice of the disentanglement weight and has the higher evidence at $L=5$ and $6$: the two strategies are points on a trade-off between interpretability and fit to the data. Notably, the latent variables for $\tau$ and $A_s$ share almost no information in any model, consistent with large-scale polarisation separating the two, as in standard analyses. A deep-learning approach can thus give a compact, data-driven parametrisation of the CMB spectra.
\end{abstract}


\section{Introduction} \label{sec:intro}

Building physically motivated cosmological models that fit the observed data is a central aim of cosmology. Traditionally, this has been done theoretically, using physical understanding of the processes that govern the evolution and structure of the components of the Universe, backed by mathematical modelling that ensures good agreement between the predicted observables and the cosmological observations \citep{Weinberg:2008zzc,Peebles:1994xt}. Among several maximally symmetric spaces that have been employed to model our universe \citep{Landau:1975pou}, the Friedmann--Lema\^{\i}tre--Robertson--Walker (FLRW) cosmology is a natural choice to describe a universe that is homogeneous and isotropic on large scales, at least from recombination to today. Within the FLRW framework, the $\Lambda$ Cold Dark Matter ($\Lambda$CDM) model is considered the benchmark of cosmology, owing to its simplicity and the extent of its agreement with observations. It describes the dynamics of the Universe, from the Big Bang to the present day, within a single framework with only six parameters, namely the physical densities of baryons ($\omega_b$) and cold dark matter ($\omega_c$), the reduced Hubble constant ($h$), the optical depth to reionisation ($\tau$), and the amplitude ($A_s$) and spectral index ($n_s$) of the primordial scalar perturbations \citep{Dodelson:2020bqr}. In that sense, our understanding of the universe depends both on the assumed background geometry (FLRW) and on the assumed model ($\Lambda$CDM).

However, modern cosmology has entered an era of unprecedented precision, driven by large volumes of high-quality data across diverse scales and epochs, and $\Lambda$CDM now faces several challenges \citep{Verde:2019ivm,DiValentino:2020zio,DiValentino:2021izs,Perivolaropoulos:2021jda,Schoneberg:2021qvd,Abdalla:2022yfr,Kamionkowski:2022pkx,Giare:2023xoc,Hu:2023jqc,Verde:2023lmm,DiValentino:2024yew,Efstathiou:2024dvn,CosmoVerseNetwork:2025alb,Riess:2021jrx,DES:2021wwk}. These challenges warrant a closer look at the assumptions that enter the analysis, with the choice of parameters being one of them. The six parameters of $\Lambda$CDM are defined on theoretical grounds and for ease of physical interpretation, and standard Markov chain Monte Carlo (MCMC) analyses typically sample directly in them. The data, however, most strongly constrain combinations of these parameters rather than the individual ones \citep{Efstathiou:1998xx,Kosowsky:2002zt}. Sampling in a rigid theoretical parametrisation could therefore lead to complex degenerate posteriors and leave the results vulnerable to prior-volume effects \citep{Trotta:2008qt,Handley:2019wlz}.

Various statistical and algorithmic strategies have been developed to address these inferential challenges. Examples include profile likelihood methods, which avoid prior-volume effects \citep{Gomez-Valent:2022hkb,Herold:2021ksg,Herold:2024enb}, and simulation-based inference, which dispenses with an explicit likelihood \citep{Alsing:2019xrx}. A further route is to extract the effective underlying dimensionality of the data directly, without imposing a physical parametrisation, using machine-learning techniques such as principal component analysis \citep{Huterer:2002hy,Leach:2005av,Peiris:2009wp,Hu:2003gh}, support vector machines \citep{Lochner:2016hbn,Jones:2016gej} and neural/deep-learning networks \citep{Charnock:2018ogm,Portillo_2020}. 

Recently, deep learning architectures that can learn effective representations directly from the data have been applied to a variety of cosmological data sets \citep{Dvorkin:2022pwo,Rojas:2025wdm,Shah:2024slr}. Instead of working within a predefined parametrisation, such an architecture can discover one that is optimised for the data, and it does not rely on any \textit{a priori} information about specific cosmological models. In particular, encoder-decoder networks such as autoencoders \citep{Hinton:2006tev} and variational autoencoders (VAE) \citep{Kingma:2013hel,Rezende:2014azh} have become popular for identifying the essential features of cosmological data sets, as well as for reconstructing them independent of a cosmological model. A VAE compresses each data vector into a small number of latent variables from which the data can be reconstructed. It also regularises the distribution of these variables, so that the latent space forms a compact and smooth parametrisation of the data. VAEs have been shown to reconstruct and generate non-linear cosmic structures \citep{Lanusse_2021}, to learn compact and semantically rich latent representations of field level cold dark matter simulation data \citep{Kannan:2025cem}, to map dark matter onto baryonic fields when trained on cosmological simulations \citep{Horowitz:2021olb}, and to encode complementary representations of baryonic effects \citep{Lin:2025ryr}.

Among all cosmological observations, the temperature and polarisation anisotropies of the cosmic microwave background (CMB) are among the most precisely measured, from the Wilkinson Microwave Anisotropy Probe (WMAP) \citep{WMAP:2012nax} through Planck 2015 \citep{Planck:2015fie} to the Planck 2018 release \citep{Planck:2018vyg}, complemented with similar missions like the South Pole Telescope (SPT) \citep{SPT-3G:2014dbx} and Atacama Cosmology Telescope (ACT) \citep{AtacamaCosmologyTelescope:2025blo}. These measurements have led to exceptionally tight constraints on the $\Lambda$CDM parameters, and CMB physics is among the most rigorously tested and theoretically best understood areas of cosmology. Six parameters are needed to fit the CMB within $\Lambda$CDM, and the data leave little room for more: additional parameters do not significantly improve the goodness of fit. Combining the CMB with other probes can make extra parameters useful, but the improved fit then comes at the price of a more complex model, so a compromise has to be struck with the help of model selection criteria \citep{Liddle:2004nh}. In the conventional model-dependent route, this exercise has to be repeated whenever a new model is introduced. A data-driven parametrisation turns it into a one-shot exercise, carried out without bias from any individual pre-defined model. In this work we follow the data-driven route and ask a more basic question: how many independent degrees of freedom do the CMB data contain if no physical parametrisation is imposed in advance? Before such a method can be trusted on extensions of the standard cosmological model, it has to be tested on a case where the expected answer can be checked. We, hence, do not claim complete theory independence, since the training spectra are generated within $\Lambda$CDM. However, no cosmological model is built into the network or its latent space, and this is what makes the outcomes data-driven. 

\citet{Piras:2025eip} recently demonstrated the efficacy of VAEs in parametrising the CMB temperature anisotropy power spectrum (TT) at $\ell\ge30$ into a disentangled, data-driven latent space. They manually tuned the weight of the Kullback--Leibler (KL) divergence term to obtain disentangled latent variables, interpreted them with mutual information (MI) \citep{cover2012elements}, found that five latent variables suffice (one fewer than the six $\Lambda$CDM parameters) to explain the data, and obtained constraints from Planck 2018 TT data in latent space. A subsequent study by \citet{Sun:2025ypd} included binned TT, TE and EE spectra at $\ell\ge30$ in a parameter-conditioned VAE and again found that five latent variables suffice. In their model, information about each parameter is spread over several latent variables, and no single latent variable corresponds cleanly to a single parameter.

Three aspects of the problem remain open. First, the earlier studies use only $\ell\ge30$ \citep{Piras:2025eip,Sun:2025ypd}. At these multipoles the spectra depend on $A_s$ and $\tau$ essentially only through the combination $A_se^{-2\tau}$, which makes these two parameters degenerate and accounts for the five latent variables found there instead of the six theoretical parameters of $\Lambda$CDM. Whether extending to $\ell\geq 2$ and including polarisation, in particular the reionisation bump \citep{Zaldarriaga:1996ke} in the large-scale EE spectrum, changes the number of latent variables the data require has, to our knowledge, not been studied with a VAE, and we do not assume the answer \textit{a priori}. Six parameters generate the spectra, but the data can be sensitive to fewer independent combinations of them, as is the case for $A_s$ and $\tau$ at $\ell\geq 30$ \citep{Planck:2018vyg}, and a VAE is free to merge such degrees of freedom. How many latent variables the full CMB data require is therefore an open question. Second, to our knowledge, a clean one-to-one correspondence between latent variables and the base $\Lambda$CDM parameters has not been obtained by tuning the KL divergence weight or by conditioning alone. Accurate reconstruction does not guarantee that each latent variable follows one physical direction, and a representation that reconstructs the spectra well can still be hard to interpret \citep{DBLP:journals/corr/abs-1811-12359}. Third, the training objective (loss function) with several terms needs a weight for each of them. These weights are usually chosen by hand and tuned over many training runs. A scheme that chooses all of them automatically from the data in a single run would remove this arbitrariness, in addition to reducing overall computational cost.

We address the first two points by asking, first, how many latent variables ($L$) suffice to fit the full CMB data and, second, to what extent these latent variables can be mapped onto the six standard $\Lambda$CDM parameters. The analysis uses the combined TT, TE and EE spectra over $2\le\ell\le2500$, in three configurations of the reconstruction loss that treat the low- and high-$\ell$ ranges and the temperature and polarisation spectra differently, so that the role of polarisation is assessed explicitly rather than being absorbed into a single data vector. The central feature of our approach is that no cosmological model or theory is built into the network architecture, i.e., the reconstruction and KL divergence terms of the loss carry no physical information, and the encoder and decoder act only on the spectra. Moreover, the latent variables are given complete freedom: the parameter each one should follow is not fixed in advance, and all share the same standard normal prior. Nor is the number of latent variables imposed strongly; we train separate models with $L=5$, $6$ and $7$, and the comparison of their fits to real data shows how many latent variables are needed. The range is centred on six, i.e. the number of $\Lambda$CDM parameters, so that the neighbourhood of the expected answer is searched. Of course, the same scan can in principle be extended to any range of $L$ until the minimum (and optimal) number is found through the statistical measures employed in this work. To encourage the latent variables to follow distinct physical directions, an explicit disentanglement term is added to the training objective. This term uses the parameters of the training samples only as a soft penalty against latent variables that mix several parameters or duplicate one another. It does not fix $L$, and it does not assign any latent variable to a given parameter. Each latent variable is still free to follow whichever parameter the data favours, and the resulting mapping to cosmological parameters is inferred only post-training. The aim is not merely accurate reconstruction, but a compact and interpretable data-driven parametrisation of the CMB spectra.

These considerations make the training objective more complex. Because physical effects and noise levels vary across angular scales, the reconstruction loss must be split across multipole ranges, each with its own hyperparameter that sets its importance in the total loss function, and the disentanglement term introduces a further hyperparameter, $w_{d}$, that sets its importance relative to the reconstruction and regularisation terms. This latter hyperparameter is hard to choose by hand: increasing $w_{d}$ improves the separation of latent information but can affect reconstruction accuracy, and the best balance may depend on the specific configuration and the multipole ranges. On the other hand, a manual scan over $w_{d}$, together with the scale-dependent reconstruction hyperparameters, would require a large number of training runs, and the result would rely on manual tuning. To avoid this, multi-task learning with uncertainty weighting \citep{Cipolla2018} is used, in which every loss term has its hyperparameter made trainable, so that it is learned together with the network in a single training run. We consider two strategies for the disentanglement weight $w_d$. In the semi-automated (SA) strategy, the weights of the reconstruction and KL divergence terms are learned, while $w_d$ is fixed by hand to a large value ($w_d=5$). In the completely automated (CA) strategy, all the weights, including $w_d$, are learned from the data in the same training run, so that no hyperparameter is chosen by hand. We use both because they answer different questions: SA shows what a strong disentanglement loss achieves, CA shows whether the weights can be set entirely from the data, and comparing them shows how the number of latent variables and the interpretability depend on the strength of the disentanglement loss. The main contributions of this work are: (i) a deep-learning parametrisation of the CMB in which no cosmological model is built into the network or its latent space, covering the full multipole range, including polarisation, (ii) an answer to the two questions posed above, namely how many latent variables suffice to fit the full CMB data and to what extent they map onto the six $\Lambda$CDM parameters, and (iii) the automated tuning of the hyperparameters of the reconstruction and KL divergence terms, and in the CA strategy also of the disentanglement term, within a single training run. The framework is implemented in our code \texttt{OCCAMS} (Optimal Components of CMB with Autoencoders and Model Selection).

The paper is organised as follows. Section~\ref{sec:vae} introduces VAEs and the basic form of their loss function. Section~\ref{sec:method} describes the network architecture, the training data, the loss function with the disentanglement term, and the multi-task tuning of the hyperparameters. Section~\ref{sec:results} presents the reconstruction accuracy, the interpretation of the latent space through normalised MI, and the latent-space fits to Planck 2018 data with the resulting model comparison metrics. We discuss the results in Section~\ref{sec:discussion} and conclude in Section~\ref{sec:conclusion}.

\section{Deep Generative Modelling and Variational Autoencoders} \label{sec:vae}

A Variational Autoencoder (VAE) \citep{Kingma:2013hel,Rezende:2014azh} learns a compact representation of its input in an unsupervised manner. An encoder network compresses each input spectrum $\bm{x}$ into a low-dimensional latent vector $\bm{z}\in\mathbb{R}^L$ (the ``bottleneck''), and a decoder network reconstructs the spectrum from $\bm{z}$. Unlike a standard autoencoder, the encoder does not output a single point. For each input it outputs a vector of means $\bm{\mu}(\bm{x})$ and a vector of standard deviations $\bm{\sigma}(\bm{x})$ of a Gaussian distribution $q_{\phi}(\bm{z}|\bm{x})$, where $\phi$ denotes the parameters of the encoder network, and $\bm{z}$ is drawn from this distribution before it is passed to the decoder. The network is trained by minimising,
\begin{equation}
\mathcal{L} = \mathcal{L}_{\text{recon}}\left(D_{\ell,\text{true}}, D_{\ell,\text{pred}}\right) + \beta\,\mathcal{D}_{KL}\big[q_{\phi}(\bm z|\bm x)\,\|\,p(\bm z)\big].
\end{equation}

The first term measures the network's reconstruction accuracy through the mean squared error (MSE) between the rescaled reconstructed angular power spectra $D_{\ell,\text{pred}}=\frac{\ell(\ell+1)}{2\pi}C_{\ell,\text{pred}}$ and the rescaled target spectra $D_{\ell,\text{true}}=\frac{\ell(\ell+1)}{2\pi}C_{\ell,\text{true}}$. Here the target ``true'' spectra are the spectra of the training set, which are also the input to the encoder, so that the network is trained to reproduce its own input. A VAE has to be trained on a large set of spectra that covers the range it is expected to represent, and since there is no analytic description of the space of possible CMB spectra, this set has to be produced by a simulator. We use the $\Lambda$CDM model purely as such a simulator of realistic spectra, because it is the standard model that fits the CMB data well. Its inputs are the parameters $\bm\Theta=(\omega_{\rm b},\,\omega_{\rm cdm},\,h,\,\tau,\,n_{\rm s},\,\ln(10^{10}A_{\rm s}))$: the physical densities of baryons and cold dark matter, the reduced Hubble constant, the optical depth to reionisation, and the spectral index and amplitude of the primordial scalar perturbations. The network, however, only ever sees the resulting spectra. It is not told how many parameters were used to generate them, nor what they are, and the number of latent variables it needs is not set by this number, since the spectra may depend on fewer independent combinations of the parameters than there are parameters. Further details about the training process are described in Section~\ref{subsec:trainingdata}. The second term is the KL divergence regulariser, which encourages the learned latent distribution $q_{\phi}(\bm{z}|\bm{x})$ to be similar to a predefined prior distribution $p(\bm{z})$. The prior is taken to be a multivariate Gaussian with zero mean and identity covariance, $p(\bm{z})=\mathcal{N}(0,I)$.

A primary challenge in training such VAEs is balancing the reconstruction accuracy against the disentanglement of the latent space. This balance is governed by the hyperparameter $\beta$ \citep{Burgess:2018hqi}, which multiplies the KL divergence term in the loss function. An optimal $\beta$ is not known in advance. It is usually found by training the model several times with different values and choosing the one that gives the best compromise between reconstruction accuracy and interpretability. The modifications to the loss function required for the objectives of this work introduce further hyperparameters of the same kind beyond $\beta$, which are discussed in the following sections. Scanning over all of them by hand is computationally prohibitive, which motivates the multi-task learning approach of Section~\ref{sec:algo}.

\section{Methodology} \label{sec:method}

\subsection{Architecture}

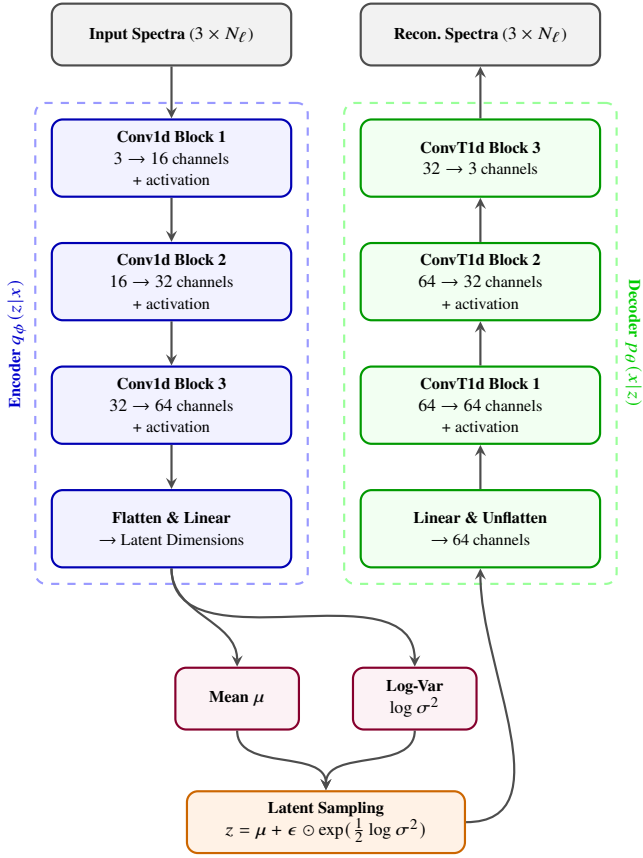
\begin{figure}
\centering
\resizebox{\columnwidth}{!}{%
\begin{tikzpicture}[
>=stealth,
layer/.style={rectangle, draw=blue!70!black, thick, rounded corners, align=center, minimum width=3.1cm, minimum height=1cm, font=\scriptsize, fill=blue!5},
declayer/.style={rectangle, draw=green!60!black, thick, rounded corners, align=center, minimum width=3.1cm, minimum height=1cm, font=\scriptsize, fill=green!5},
data/.style={rectangle, draw=black!70, thick, rounded corners, align=center, minimum width=3.1cm, minimum height=0.8cm, font=\scriptsize, fill=gray!10},
latentbox/.style={rectangle, draw=purple!70!black, thick, rounded corners, align=center, minimum width=1.6cm, minimum height=0.8cm, font=\scriptsize, fill=purple!5},
arrow/.style={->, thick, black!70}
]
\node[data] (input) at (0, 0) {\textbf{Input Spectra} ($3 \times N_\ell$)};
\node[layer] (enc1) at (0, -1.6) {\textbf{Conv1d Block 1} \\ $3 \to 16$ channels \\ + activation};
\node[layer] (enc2) at (0, -3.2) {\textbf{Conv1d Block 2} \\ $16 \to 32$ channels \\ + activation};
\node[layer] (enc3) at (0, -4.8) {\textbf{Conv1d Block 3} \\ $32 \to 64$ channels \\ + activation};
\node[layer] (fc_enc) at (0, -6.4) {\textbf{Flatten \& Linear} \\ $\to$ Latent Dimensions};
\node[latentbox] (mu) at (0.85, -8.6) {\textbf{Mean} $\mu$};
\node[latentbox] (logvar) at (3.15, -8.6) {\textbf{Log-Var} \\ $\log\sigma^2$};
\node[latentbox, fill=orange!10, draw=orange!80!black, minimum width=3.6cm] (z) at (2.0, -10.2) {\textbf{Latent Sampling} \\ $z = \mu + \epsilon \odot \exp(\frac{1}{2}\log\sigma^2)$};
\node[declayer] (fc_dec) at (4.0, -6.4) {\textbf{Linear \& Unflatten} \\ $\to$ 64 channels};
\node[declayer] (dec1) at (4.0, -4.8) {\textbf{ConvT1d Block 1} \\ $64 \to 64$ channels \\ + activation};
\node[declayer] (dec2) at (4.0, -3.2) {\textbf{ConvT1d Block 2} \\ $64 \to 32$ channels \\ + activation};
\node[declayer] (dec3) at (4.0, -1.6) {\textbf{ConvT1d Block 3} \\ $32 \to 3$ channels};
\node[data] (output) at (4.0, 0) {\textbf{Recon. Spectra} ($3 \times N_\ell$)};
\draw[arrow] (input) -- (enc1);
\draw[arrow] (enc1) -- (enc2);
\draw[arrow] (enc2) -- (enc3);
\draw[arrow] (enc3) -- (fc_enc);
\draw[arrow] (fc_enc.south) to[out=270, in=90] (mu.north);
\draw[arrow] (fc_enc.south) to[out=270, in=90] (logvar.north);
\draw[arrow] (mu.south) to[out=270, in=90] (z.north);
\draw[arrow] (logvar.south) to[out=270, in=90] (z.north);
\draw[arrow] (z.east) to[out=0, in=270] (fc_dec.south);
\draw[arrow] (fc_dec) -- (dec1);
\draw[arrow] (dec1) -- (dec2);
\draw[arrow] (dec2) -- (dec3);
\draw[arrow] (dec3) -- (output);
\begin{scope}[on background layer]
\node[draw=blue!40, dashed, thick, rounded corners, fit=(enc1) (fc_enc), inner sep=0.2cm, label={[rotate=90, anchor=south, font=\bfseries\scriptsize, text=blue!80!black]west:Encoder $q_\phi(z|x)$}] {};
\node[draw=green!40, dashed, thick, rounded corners, fit=(fc_dec) (dec3), inner sep=0.2cm, label={[rotate=-90, anchor=south, font=\bfseries\scriptsize, text=green!80!black]east:Decoder $p_\theta(x|z)$}] {};
\end{scope}
\end{tikzpicture}}
\caption{The VAE network architecture. The input spectra are compressed via successive 1D convolutions into a latent distribution defined by $\mu$ and $\log\sigma^2$. A stochastic sample $z$ is drawn and subsequently upsampled via transposed convolutions to reconstruct the original spectra.\vspace{-0.4cm}}
\label{fig:cvae_architecture}
\end{figure}

The compression and reconstruction pipeline is a 1D convolutional VAE, designed to map the high-dimensional CMB spectra into a continuous, low-dimensional probabilistic space. The architecture consists of three primary components: an encoder network, a stochastic latent bottleneck, and a decoder network (illustrated in Fig.~\ref{fig:cvae_architecture}). The full pipeline is implemented in \texttt{OCCAMS} (Optimal Components of CMB with Autoencoders and Model Selection), which is our code developed for this analysis.

Throughout the network we use the trainable activation function adopted by \citet{Piras:2025eip}. For an input $x$ it is defined as, $f(x) = x \cdot \left[ \gamma + (1 - \gamma)\,\mathrm{sigmoid}(\kappa x) \right]$, where $\gamma$ and $\kappa$ are learnable parameters optimised during training together with the other parameters of the network. It adds a trainable non-linearity that interpolates between a linear and a sigmoid-gated response. Each convolutional layer, except the final output layer of the decoder, is followed by this activation and 1D batch normalisation to stabilise training. 

The encoder $q_\phi(z|x)$ serves as the inference network, compressing the multi-channel input sequences into the parameters of the approximate posterior distribution. The input consists of three channels, the spectra $D_\ell^{TT}$, $D_\ell^{TE}$ and $D_\ell^{EE}$, each sampled at $N_{\ell}$ multipoles over $2\le\ell\le2500$. The network applies three successive 1D convolutional blocks. The first block expands the 3 input channels to 16 channels using a stride of 3 and a kernel size chosen so that the decoder reproduces exactly $N_\ell$ multipoles. The second block expands the 16 channels to 32 channels using a kernel size of 16 and a stride of 3. The third block further expands the 32 channels to 64 channels using a kernel size of 3 and a stride of 3. Following the feature extraction, the 64-channel output is flattened into a 5760-dimensional vector. A fully connected linear layer then projects this representation into a 2$L$-dimensional vector, which is split equally to parametrise the mean ($\mu$) and the log-variance ($\log\sigma^2$) of the latent distribution. To ensure numerical stability during optimisation, the log-variance is clamped between $[-10, 10]$.

The latent representation is an $L$-dimensional Gaussian distribution with diagonal covariance. $L$ is a fixed choice of the architecture for each model: we train separate models with $L=5$, 6 and 7 and compare them in Section~\ref{sec:post}. This range is chosen around the six parameters of $\Lambda$CDM, and can be extended to any range of $L$ if the scan does not enclose the minimum number. To allow gradients to backpropagate through the stochastic sampling process, the standard reparametrisation trick is employed: the latent vector is sampled as $z = \mu + \epsilon \odot \exp(\tfrac{1}{2}\log\sigma^2)$, where $\epsilon \sim \mathcal{N}(0, I)$ is standard Gaussian noise and $\odot$ represents element-wise multiplication.

The decoder $p_\theta(x|z)$, where $\theta$ denotes the parameters of the decoder network, acts as the generative network, mapping the $L$-dimensional latent vector $z$ back to the space of the three spectra. The sampled vector $z$ is first passed through a dense layer, expanding it to 5760 features. This is reshaped into a 3D tensor of 64 channels with a spatial length of 90. The spatial resolution is progressively upsampled using a sequence of 1D transposed convolutions (\texttt{ConvTranspose1d}). The first block maintains 64 channels using a kernel size of 3 and a stride of 3, accompanied by the activation and batch normalisation. The second block reduces the feature maps from 64 to 32 channels using a kernel size of 16 and a stride of 3. Finally, the third block projects the 32 channels back to the original 3 output channels ($D_\ell^{TT}$, $D_\ell^{TE}$, $D_\ell^{EE}$) using a stride of 3 and a kernel size matching that of the first encoder block. No activation function is applied at this final layer, allowing the network to directly output the reconstructed values.

\subsection{Training Data} \label{subsec:trainingdata}

A total of 200,000 sets of spectra ($C_\ell^{TT}$, $C_\ell^{TE}$, $C_\ell^{EE}$) are generated with the Einstein--Boltzmann solver \texttt{CLASS} \citep{Lesgourgues:2011re,Blas:2011rf}. \texttt{CLASS} numerically solves the coupled equations that govern the evolution of the photons, baryons, dark matter and neutrinos in the early Universe, and from their solution computes the CMB angular power spectra. For the purposes of this work it can be regarded as a deterministic simulator that maps a set of input cosmological parameters (six in the case of $\Lambda$CDM) to the corresponding temperature and polarisation spectra, and it is used only to produce the training examples. The six $\Lambda$CDM parameters are drawn with Latin hypercube sampling within the ranges given in Table~\ref{tab:prior_ranges}. In this scheme the range of each parameter is divided into as many equal intervals as there are samples, one value is drawn at random within each interval, and the values of the different parameters are combined in a random order. Every parameter range is thus covered evenly, while the combinations of parameters are random. The training set therefore contains no preferred cosmology and no imposed correlations between the parameters, so any dependence of the spectra on fewer than six independent combinations of them is a property of the spectra and not of how the sample was drawn. The spectra cover $2\le\ell\le2500$. The samples are split into training ($80\%$), validation ($10\%$) and test ($10\%$) sets.

The spectra are converted to $D_\ell=\ell(\ell+1)C_\ell/2\pi$ and pre-processed following the scheme of \citet{Piras:2025eip}. Each spectrum is first divided by a arbitrary reference spectrum (different for TT, TE and EE), which facilitates training. A base-10 logarithm is then taken, followed by normalisation of the resulting data vectors. Since the TE spectra change sign, a vertical offset is imposed so that the logarithms are well defined. The VAE output is processed through the same steps in reverse order to obtain $D_{\ell,\rm pred}$.

\begin{table}
\centering
\renewcommand{\arraystretch}{1.2}
\begin{tabular}{c c}
\toprule
\textbf{Parameter} & \textbf{Sampling Range} \\
\midrule
$\omega_{\mathrm{b}}$ & $[0.020, 0.024]$ \\
$\omega_{\mathrm{cdm}}$ & $[0.10, 0.13]$ \\
$h$ & $[0.62, 0.80]$ \\
$\tau$ & $[0.01, 0.13]$ \\
$n_{\mathrm{s}}$ & $[0.92, 1.01]$ \\
$\ln(10^{10}A_{\mathrm{s}})$ & $[2.90, 3.18]$ \\
\bottomrule
\end{tabular}
\caption{Uniform sampling ranges used to generate the training data set via a Latin hypercube. Following \citet{Piras:2025eip}, the ranges span approximately $10\sigma$ around the \textit{Planck} 2018 baseline best-fit values. There are two exceptions to these bounds: the lower limit for the optical depth ($\tau$) is restricted to $0.01$ to prevent non-physical negative values, and the upper limit for the reduced Hubble constant ($h$) is broadened to encompass the local distance-ladder measurement from the SH0ES collaboration \citep{Riess:2021jrx}.}
\label{tab:prior_ranges}
\end{table}

\subsection{Loss Function}

\subsubsection{Temperature Anisotropies (TT)}
\label{subsec:TTtheory}

The temperature fluctuations of the CMB, $\Delta T(\hat{n})/T = \sum_{\ell, m} a_{\ell m} Y_{\ell m}(\hat{n})$, are described statistically by the angular power spectrum $C_\ell^{TT} = \langle |a_{\ell m}|^2 \rangle$. Its shape changes with angular scale, because different physical processes dominate at different scales \citep{Hu:2001bc,Zaldarriaga:1997ch}. The positions of the acoustic peaks give the angular size of the sound horizon $\theta_s$, the relative height of the odd and even peaks is sensitive to $\omega_{\rm b}$, the peak heights to $\omega_{\rm cdm}$, the spectral index $n_s$ tilts the spectrum as a whole, and the overall amplitude is set by $A_s$. At the largest scales, cosmic variance limits how well parameters can be measured \citep{Planck:2018vyg}. 

\citet{Piras:2025eip} trained a VAE on $\Lambda$CDM TT spectra at $\ell\ge30$. To find the optimal number of latent variables, they increased the latent dimension step by step, and kept the smallest number for which the reconstruction was well within the Planck $1\sigma$ error. With four latent variables the reconstruction error became comparable to the Planck error, while five were found to be enough. The latent variables were compared with the cosmological parameters only after training, using MI and latent traversals. They correspond to the overall amplitude, the positions of the acoustic peaks, the peak heights together with the spectral tilt, the even-odd modulation of the peaks, and the smoothing of the peaks by gravitational lensing.

The result has two implications. First, the TT spectrum at $\ell\ge30$ has five independent degrees of freedom, one fewer than the six parameters of $\Lambda$CDM, and this number was found from the VAE analysis and not set in advance. Second, the missing degree of freedom is related to the degeneracy between $A_s$ and $\tau$ discussed in Section~\ref{subsec:pol}: the amplitude latent variable follows the combination $A_s e^{-2\tau}$, and not $A_s$ and $\tau$ separately.

In \citet{Piras:2025eip}, the weight of the KL divergence term in the loss was manually tuned by the authors for each model to obtain disentangled latent variables, using the MI between the latent variables as the guide. In this work the weights of the reconstruction and KL divergence terms, and in the CA strategy also of the disentanglement term, are instead learned automatically in a single training run, which removes the need to tune these weights by hand.

\subsubsection{Including Polarisation (TE, EE)} \label{subsec:pol}

\begin{table}
\centering
\renewcommand{\arraystretch}{1.5}
\begin{tabular}{c c}
\toprule
 \textbf{Configuration Name} & \textbf{Reconstruction Terms $\mathcal{T}$} \\
\midrule
 lowTTTEEE+full & $\mathcal{T}=\{\mathcal{L}_{\text{low}\,\ell},\ \mathcal{L}_{\text{full}\,\ell}\}$  \\[1ex]
 lowTTTEEE+high & $\mathcal{T}=\{\mathcal{L}_{\text{low}\,\ell},\ \mathcal{L}_{\text{high}\,\ell}\}$  \\[1ex]
 lowTT-TEEE+high & $\mathcal{T}=\{\mathcal{L}_{\text{lowTT}\,\ell},\ \mathcal{L}_{\text{lowTEEE}\,\ell},\ \mathcal{L}_{\text{high}\,\ell}\}$  \\
\bottomrule
\end{tabular}
\caption{Summary of the three loss function configurations considered here. }
\label{tab:loss_configurations}
\end{table}

The temperature spectrum alone, however, has a built-in degeneracy. Thomson scattering off free electrons after reionisation reduces the power at small scales ($\ell \gtrsim 30$) by a factor of $e^{-2\tau}$, where $\tau$ is the optical depth. As a result, across the acoustic peaks and the damping tail, $C_\ell^{TT}$ depends on $A_s$ and $\tau$ essentially only through the combination $A_s e^{-2\tau}$ \citep{Planck:2018vyg}. Because the large scales are limited by cosmic variance, the temperature spectrum alone can constrain $A_s$ and $\tau$ separately only weakly. A VAE trained on the temperature spectrum at $\ell\ge30$ can therefore resolve only this combination, and not the two parameters separately.

Polarisation at large scales carries independent information on $\tau$. Thomson scattering of the local quadrupole at reionisation produces a peak at low multipoles in the $EE$ spectrum (the ``reionisation bump''), and a corresponding feature in $TE$. Its amplitude scales approximately as $C_{\ell}^{EE}\propto A_s\tau^2$ for $\ell\lesssim 15$, so it constrains $\tau$ separately from the high-$\ell$ amplitude $A_s e^{-2\tau}$. To allow the VAE to fit both the low-$\ell$ and the high-$\ell$ regimes, we split the reconstruction loss $\mathcal{L}_{\text{recon}}\big(D_{\ell,\text{true}}, D_{\ell,\text{pred}}\big)$ into MSEs over separate multipole ranges: $\mathcal{L}_{\text{low}\,\ell}$ for $\ell\le14$, $\mathcal{L}_{\text{high}\,\ell}$ for $\ell\ge15$, and $\mathcal{L}_{\text{full}\,\ell}$ for the full range $2\le\ell\le2500$. The boundary at $\ell=14$ is chosen to isolate the multipoles $\ell\lesssim15$ at which the reionisation bump dominates, so that the low-$\ell$ polarisation, which carries the information on $\tau$, is not outweighed by the far larger number of high-$\ell$ multipoles in the loss. Unless stated otherwise, each of these is computed over the TT, TE and EE spectra.

\begin{figure*}
\centering
\begin{tikzpicture}[
>=stealth,
node distance=1cm and 0.6cm, 
data/.style={rectangle, draw=black!70, thick, rounded corners, align=center, minimum width=2.4cm, minimum height=1.1cm, font=\scriptsize, fill=green!5},
net/.style={rectangle, draw=blue!70!black, thick, rounded corners=8pt, align=center, minimum width=2.2cm, minimum height=1.2cm, font=\scriptsize, fill=blue!5},
latent/.style={rectangle, draw=purple!70!black, thick, rounded corners, align=center, minimum width=2.6cm, minimum height=1.2cm, font=\scriptsize, fill=purple!5},
lossblock/.style={rectangle, draw=red!60!black, thick, align=center, minimum width=5.5cm, minimum height=1.2cm, font=\scriptsize, fill=red!5},
mtl/.style={rectangle, draw=black, thick, align=center, minimum width=6.5cm, minimum height=1.2cm, font=\scriptsize, fill=gray!10},
arrow/.style={thick, -{Stealth[scale=1.1]}},
dashed_arrow/.style={thick, dashed, black!60, -{Stealth[scale=1.1]}},
backprop/.style={thick, red!80!black, -{Stealth[scale=1.1]}}
]
\node (enc) [net] {\textbf{Encoder} \\ $q_\phi(z|x)$};
\node (latent) [latent, right=of enc] {\textbf{Latent Space} ($\bm{z}$)};
\node (dec) [net, right=of latent] {\textbf{Decoder} \\ $p_\theta(x|z)$};
\node (output) [data, right=of dec] {\textbf{Recon. Spectra} \\ $\hat{D}_\ell^{TT}$, $\hat{D}_\ell^{TE}$, $\hat{D}_\ell^{EE}$};
\draw [arrow] (enc) -- (latent);
\draw [arrow] (latent) -- (dec);
\draw [arrow] (dec) -- (output);
\begin{scope}[on background layer]
\node[draw=blue!40, dashed, thick, rounded corners, fit=(enc) (latent) (dec) (output), inner sep=0.3cm, label={[anchor=south, font=\scriptsize, text=blue!80!black]south:Variational Autoencoder (VAE)}] (vae_box) {};
\end{scope}
\node (input) [data, above=1.2cm of enc] {\textbf{Target Spectra} \\ $D_\ell^{TT}(\Theta)$, $D_\ell^{TE}(\Theta)$, $D_\ell^{EE}(\Theta)$};
\node (class) [data, fill=orange!10, draw=orange!80!black, left=of input] {\textbf{Theory Gen.} \\ $\Lambda$CDM (\texttt{CLASS})};
\draw [arrow] (class) -- (input);
\draw [arrow] (input) -- node[right, font=\scriptsize] {Input} (enc);
\node (losses) [lossblock, below=1.5cm of latent] {\textbf{Task Loss Computation} \\ $\mathcal{L}_{\text{total}} = 
\sum_{i\in\mathcal{T}} w_i\mathcal{L}_i
+ \beta D_{KL} + w_{d}\mathcal{L}_{\text{dis}}$};
\node (mtl) [mtl, below=0.6cm of losses] {\textbf{Multi-Task Uncertainty Weighting} \\ $\mathcal{L}_{\text{total}} = \sum_{i} \left( \frac{\mathcal{L}_i}{2\sigma_i^2} + \log \sigma_i \right)$};
\draw [arrow] (losses) -- (mtl);
\begin{scope}[on background layer]
\node[draw=red!30, dashed, thick, rounded corners, fit=(losses) (mtl), inner sep=0.3cm, label={[anchor=south, font=\scriptsize, text=red!80!black]south:Single-Run Automated Weighting}] {};
\end{scope}
\draw [dashed_arrow] (output.south) to[out=270, in=0] node[pos=0.2, right, font=\scriptsize] {Pred $\hat{D}_\ell$} (losses.east);
\draw [dashed_arrow] (latent.south) -- node[right, font=\scriptsize] {} (losses.north);
\draw [backprop] (mtl.west) to[out=180, in=270] node[pos=0.20, left, font=\scriptsize, align=center] {\textbf{Backprop} \\ Update $\phi, \sigma_i$} (enc.south);
\draw [backprop] (mtl.east) to[out=0, in=270] node[pos=0.20, right, font=\scriptsize, align=center] {Update $\theta, \sigma_i$} (dec.south);
\end{tikzpicture}
\caption{The end-to-end Multi-Task Learning VAE pipeline for CMB parametrisation, shown for the CA strategy. Top: The theoretical $\Lambda$CDM model is simulated via \texttt{CLASS} to generate target temperature and polarisation spectra. Middle (violet dashed box): These spectra are compressed by the encoder into an $L$-dimensional probabilistic latent space, before being reconstructed by the decoder. Bottom (red dashed box): The extraction of data to compute the individual loss terms during training. A trainable parameter $\sigma_i$ is attached to each term. Backpropagation (solid red curves) updates the network parameters ($\phi, \theta$) and the $\sigma_i$ together, so that the loss weights are set in a single training run. In SA strategy, $w_d$ is fixed and the disentanglement loss term is not assigned any $\sigma_i$.}
\label{fig:full_pipeline}
\end{figure*}
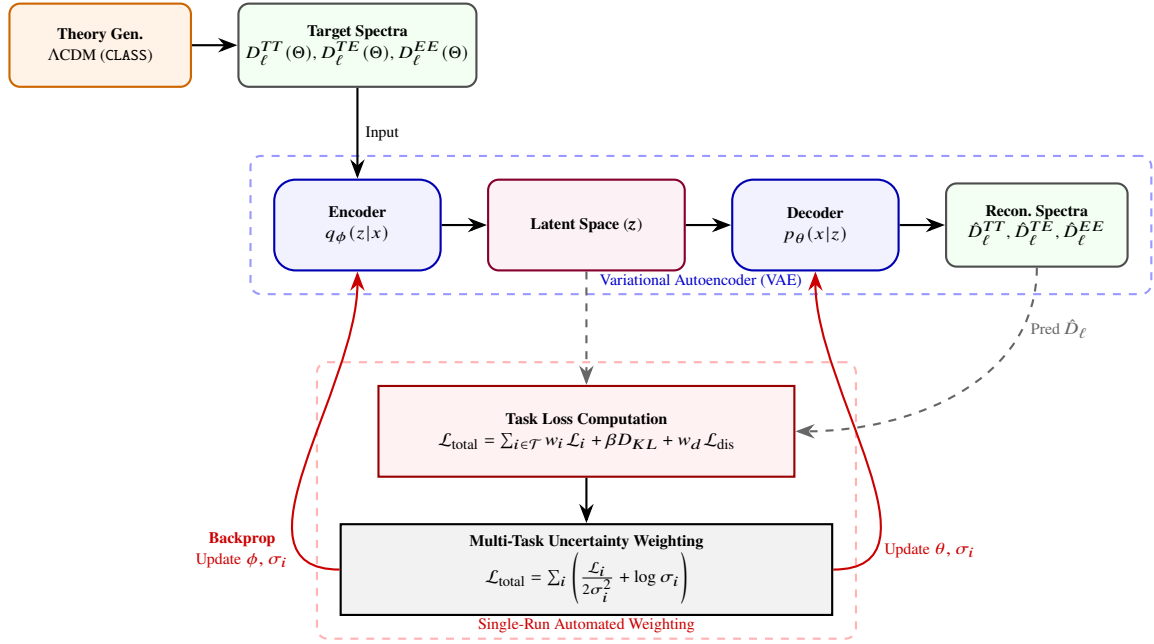

The reconstruction loss is built from these terms in three different combinations, which define the three configurations studied in this work. All three use the same training spectra and differ only in how the reconstruction error is grouped and weighted. The total loss function has the general form,
\begin{equation}
\mathcal{L}_{\text{total}} = \sum_{i\in\mathcal{T}} w_i\,\mathcal{L}_i
+ \beta\,\mathcal{D}_{KL} + w_{d}\,\mathcal{L}_{\text{dis}},
\label{eq:generic_loss}
\end{equation}
where $\mathcal{T}$ is the set of reconstruction terms, which differs between configurations, and $w_i$, $\beta$ and $w_d$ are the loss weights, i.e. the hyperparameters that set the importance of each term. The three configurations were chosen to separate the roles of the temperature and polarisation data and of the low- and high-$\ell$ ranges: the first two differ in whether the high-$\ell$ error is weighted separately from, or together with, the low-$\ell$ error, and the third weights the low-$\ell$ temperature and polarisation errors separately. They are summarised in Table~\ref{tab:loss_configurations}. Here $\mathcal{L}_{\text{lowTT}\,\ell}$ is the MSE of the TT spectrum for $\ell\le14$, and $\mathcal{L}_{\text{lowTEEE}\,\ell}$ is the MSE of the TE and EE spectra for $\ell\le14$. The third configuration therefore gives separate loss weights to the low-$\ell$ temperature and polarisation errors. In the lowTTTEEE+full configuration, the low-$\ell$ range appears in both $\mathcal{L}_{\text{low}\,\ell}$ and $\mathcal{L}_{\text{full}\,\ell}$. Additionally, $\mathcal{L}_{\text{dis}}$ is the differentiable disentanglement loss function, described in Section~\ref{sec:dis_loss}.

\subsubsection{Differentiable Disentanglement Loss Formulation} \label{sec:dis_loss}

The reconstruction and KL terms alone do not tie the latent variables to physical quantities. A VAE trained only on them can describe the spectra equally well after its latent coordinates are mixed, and hence physical disentanglement cannot be determined without extra information. For our application, this matters for two reasons. First, in order to interpret the latent variables we need to know which physical effects each of them encodes, and to what extent. This can be assessed in a straightforward manner only if each latent variable can be compared with a physical parameter. Second, if the latent axes correspond to $\Lambda$CDM parameters, the latent posteriors can be read directly, rather than only through the decoder.

Therefore a penalty term is added to the loss function with two aims: each latent should depend on as few of the six cosmological parameters $\bm\Theta$ as possible, and different latents should depend on different parameters. We do not demand a strict one-to-one mapping, because the spectra constrain some combinations of parameters, such as $A_s e^{-2\tau}$, better than the individual parameters, so one latent may reasonably follow two of them. The loss therefore allows each latent to carry up to two parameters (top-2 purity) and favours a strict one-to-one mapping only weakly. It is computed on each batch, from the latents and the cosmological parameters that generated the training samples, using only differentiable operations, so that it can be minimised together with the reconstruction error. This term is the only place where these parameters enter the training, and it acts as a soft penalty: it does not set $L$ and does not assign any latent variable to a given parameter. The details are described in Appendix \ref{app:L_dis} and Fig.~\ref{fig:matrix_multiplication}. Note that the penalty uses the squared Pearson correlation, which measures linear dependence, whereas the normalised MI used later to evaluate the latent variables is not restricted to linear dependence.

\subsection{Algorithm: Automated Hyperparameter Tuning} \label{sec:algo}

The training objective contains several terms: the reconstruction errors on different multipole ranges, the KL divergence and the disentanglement loss. The loss weights that set their relative importance must be chosen before training. A good choice, however, is not known in advance, and finding it by trial and error would need many training runs, which is impractical given the number of loss weights and the cost of each run. To avoid this, we adapt the multi-task learning framework of \citet{Cipolla2018}, which dynamically tunes the loss weights during a single training run.

We treat each term $\mathcal{L}_i$ as a separate task and attach to it a trainable parameter $\sigma_i>0$,
\begin{equation}
\mathcal{L}_{\text{total}} = \sum_{i} \left( \frac{\mathcal{L}_i}{2\sigma_i^2} + \log \sigma_i \right),
\label{eq:mtl}
\end{equation}
where the $\sigma_i$ are optimised by gradient descent together with the network parameters. The loss weight of term $i$ is $w_i = 1/(2\sigma_i^2)$, so a larger $\sigma_i$ means a smaller loss weight. In the notation of equation~\eqref{eq:generic_loss}, the KL term has $\beta = 1/(2\sigma_{\rm KL}^2)$, and in the CA strategy of Section~\ref{sec:sa_ca} the disentanglement term has $w_d = 1/(2\sigma_d^2)$. The $\log\sigma_i$ term prevents the trivial solution $\sigma_i\to\infty$, which would set all loss weights to zero. For fixed network parameters, equation~\eqref{eq:mtl} is minimised at $\sigma_i^2=\mathcal{L}_i$, i.e.\ $w_i = 1/(2\mathcal{L}_i)$. The loss weights are therefore not free hyperparameters with a single optimal value. Rather they follow the current size of each term, so that a term whose loss stays large gets a smaller weight than one that decreases quickly. Consequently, the weight of a term grows as its loss falls. The factor $1/(2\sigma^2)$ follows from a Gaussian likelihood for regression tasks \citep{Cipolla2018}; applying it to the KL divergence and disentanglement terms is a heuristic extension. Fig.~\ref{fig:full_pipeline} shows the full algorithm.

\begin{figure*}
\centering
\includegraphics[width=\textwidth]{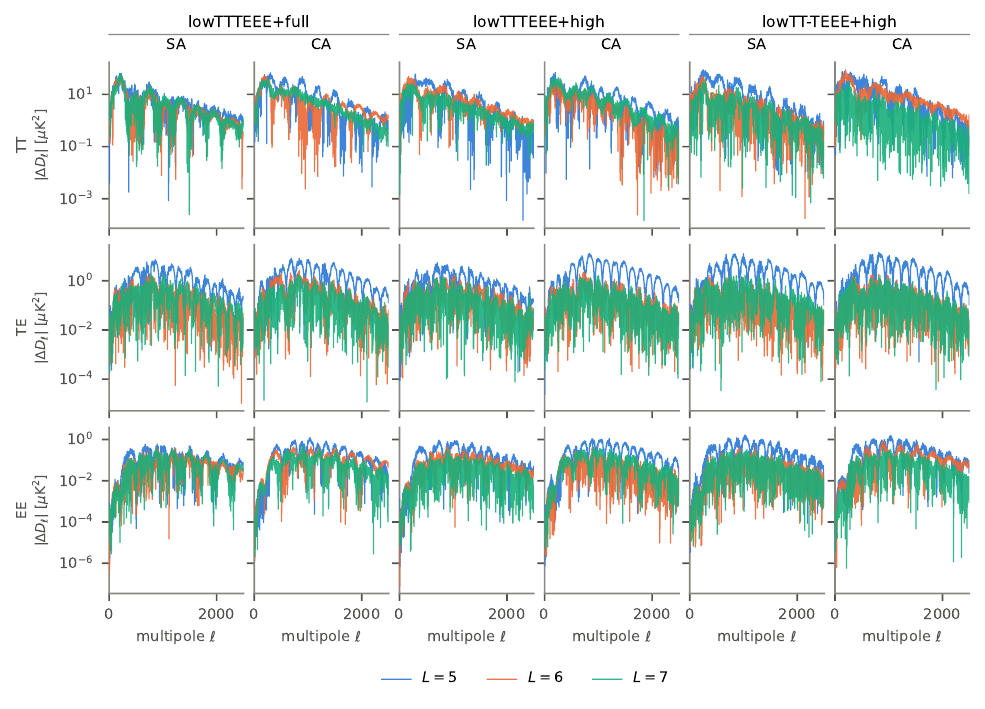}
\caption{Absolute reconstruction residuals $|\Delta D_\ell|=|D_\ell^{\rm VAE}-D_\ell^{\rm true}|$ of the Planck 2018 best-fit $\Lambda$CDM TT (top), TE (middle) and EE (bottom) spectra. Each pair of columns shows one configuration, with SA (left) and CA (right). The latent dimensions are $L=5$ (blue), $L=6$ (orange) and $L=7$ (green).}
\label{fig:recon_resid}
\end{figure*}

\subsubsection{Treatment of the Disentanglement Loss Weight} \label{sec:sa_ca}

The disentanglement loss $\mathcal{L}_{\text{dis}}$ is the only term that uses information about the cosmological parameters, through the parameter values of the training samples. It acts as a soft penalty that encourages each latent variable to follow separate features that are independent of what other latent variables follow. It does not fix the number of latent variables, it does not assign a latent variable to a given parameter, and the ordering of the latent variables remains arbitrary (Section~\ref{sec:mi}). All other terms of the loss carry no physical information.

This makes the choice of its loss weight, $w_{d}$, less straightforward than for the other terms, since $w_d$ sets how strongly the latent space is pulled towards the six cosmological parameters. To test the role of the disentanglement loss, the two strategies are implemented in equation~\eqref{eq:generic_loss} as follows:
\begin{enumerate}
\item In the \emph{semi-automated} (SA) strategy, $w_i$ and $\beta$ are learned in the same way, but $w_{d}$ is held fixed at $5.0$, and the disentanglement term is excluded from the uncertainty weighting.
\item In the \emph{completely automated} (CA) strategy, every term, including the disentanglement loss, is given its own $\sigma_i$ as in equation~\eqref{eq:mtl}, so that $w_i$, $\beta$ and $w_{d}$ are all learned in a single training run.
\end{enumerate}
The disentanglement loss therefore acts with a fixed strength in SA and with a learned strength in CA strategies.

\section{Results} \label{sec:results}

Each VAE is trained for 200 epochs, by which stage the training losses have converged to their minimum values. Combining three configurations (lowTTTEEE+full, lowTTTEEE+high, and lowTT-TEEE+high), two strategies for the disentanglement loss weight (SA and CA), and three latent dimensions, $L=5,6,7$, gives a total of 18 models. For each run, the architecture, the training settings and, for SA, the value of $w_{d}$ are fixed before training and are not tuned manually. The learned loss weights of every trained model are listed in Appendix~\ref{app:weights}. Each model is evaluated in three ways: the accuracy of the reconstructed spectra (Section~\ref{sec:recon}), the correspondence between latent variables and cosmological parameters (Section~\ref{sec:mi}), and the fit of the latent-space model to Planck 2018 data (Section~\ref{sec:post}), where the models are compared using $\hat{\chi}^2$, Akaike Information Criterion (AIC), Bayesian Information Criterion (BIC) and Bayesian evidence.

\subsection{Reconstruction}\label{sec:recon}

Fig.~\ref{fig:recon_resid} shows the absolute residuals between the spectra reconstructed by each of the 18 models and the Planck 2018 best-fit $\Lambda$CDM spectra. The trained VAE models reconstruct the CMB power spectra with absolute reconstruction residual ($|\Delta D_\ell|$) values well within the Planck 2018 uncertainties. Three features are observed across the models:
\begin{enumerate}
\item The residuals for $L=5$ (blue) lie above those for $L=6$ (orange) and $L=7$ (green) in most of the cases, so the reconstruction is generally less accurate with $L=5$ than with $L\geq6$.
\item The residuals for $L=6$ and $L=7$ are similar, so the seventh latent variable does not improve the reconstruction significantly, despite the extra degree of freedom and the higher model complexity.
\item The residuals are largest at the peaks and troughs of the spectra.
\end{enumerate}

\subsection{Interpreting the Latent Space} \label{sec:mi}

We use the normalised MI to determine how much information each latent variable shares with each $\Lambda$CDM parameter. MI is a statistical measure of the information shared between two variables \citep{Vergara_2013}. It is not restricted to linear dependence, which suits latent variables that may depend non-linearly on the parameters, and has been used to interpret latent spaces in cosmology \citep{Lucie-Smith:2022uvv,Lucie-Smith:2024xsx,Guo:2024had}. The raw MI values are computed with the \texttt{GMM-MI} package \citep{Piras:2022wso}, which fits a Gaussian mixture model to the joint distribution of the two variables, as also done by \citet{Piras:2025eip}.

Raw MI values are in units of nats and, unlike a correlation coefficient, have no fixed upper limit. We therefore divide each raw MI value by the maximum MI that can be reached for the given sample size ($N=200$ samples in our case), which bounds the normalised MI between 0 and 1. The same constant is applied to every entry of every model, so the normalisation rescales all values together and does not change their ordering. Our values are therefore not on the same scale as raw MI values in nats, and they should not be compared directly with those of \citet{Piras:2025eip}. Larger values mean that a latent variable shares more information with the corresponding cosmological parameter(s). Hereafter, MI refers to the normalised MI.

\begin{figure*}
\centering
\includegraphics[width=\textwidth]{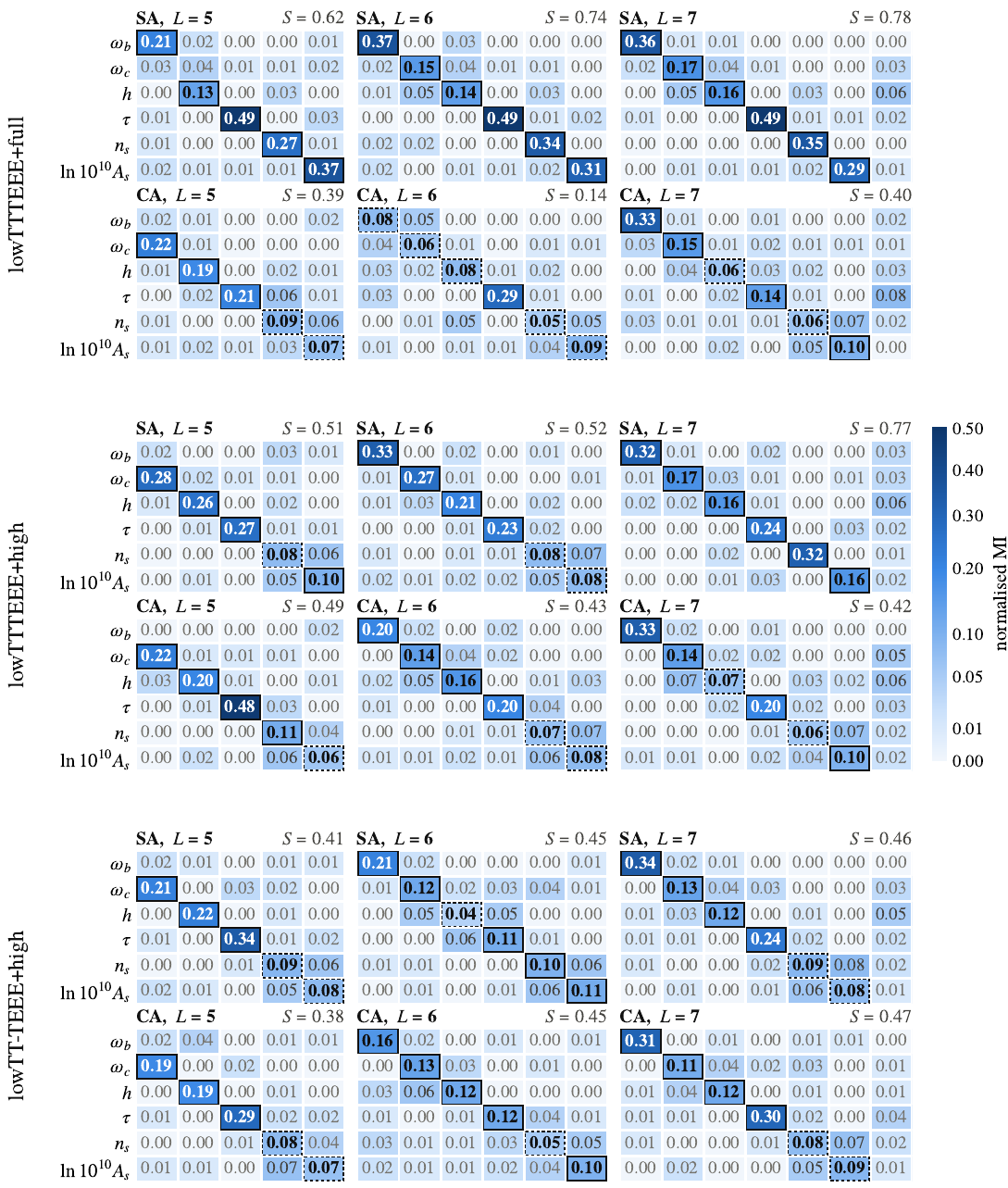}
\caption{Normalised MI between each $\Lambda$CDM parameter (rows) and each latent variable (columns) for all 18 trained models. The three blocks are the configurations, each shown for the SA and CA strategies, and the columns of panels correspond to $L=5,6,7$. The number above each panel is the disentanglement score $S$. Solid outline: parameter-latent pair assigned by the matching, with normalised MI of at least 0.10; dashed outline: assigned pair below the threshold of 0.10, counted as not captured. All values are on a common colour scale that follows a power law with exponent 0.6 so that small values can be told apart.}
\label{fig:mi_full}
\end{figure*}

To summarise the MI matrix of each model with a single number, we define a disentanglement \textit{score}. It measures both the strength and the exclusivity of the association between latent variables and cosmological parameters, and it penalises information spread over several latent variables, and cosmological parameters left without a latent variable. For a normalised MI matrix $M\in\mathbb{R}^{N\times L}$, with $N$ parameters and $L$ latent variables, each parameter is first matched to one latent variable with the Hungarian linear sum assignment algorithm \citep{https://doi.org/10.1002/nav.3800020109}, which maximises the total MI of the matched pairs. A matched pair is kept only if its MI is at least a threshold $M_{\rm min}=0.10$, chosen to separate clear associations from weak ones, and pairs below it are discarded. The parameter coverage $C = N_{\text{assigned}} / N$ is the fraction of parameters that are matched to a latent variable. For $L=5$ there are fewer latent variables than parameters, so $C\leq5/6$, which lowers the maximum possible score and should be kept in mind when comparing scores across $L$. The dependence of the score on the choice $M_{\rm min}=0.10$ is demonstrated in Appendix~\ref{app:threshold}.

For each retained pair $(i, j)$ of parameter $i$ and latent variable $j$, the quality of the association is evaluated through three quantities: purity, row redundancy and column redundancy. The purity,
\begin{equation}
P_{ij} = \frac{M_{i,j}}{\sum_{k} M_{k,j}},
\end{equation}
is the fraction of the total MI of latent variable $j$ that is due to parameter $i$. Information of parameter $i$ spread over other latent variables is penalised by a row factor,
\begin{equation}
F^{\text{row}}_i = \frac{1}{1 + \sum_{k \neq j} M_{i,k}}, \label{eq:Frow}
\end{equation}
and a latent variable $j$ that encodes other parameters as well is penalised by a column factor,
\begin{equation}
F^{\text{col}}_j = \frac{1}{1 + \sum_{k \neq i} M_{k,j}}. \label{eq:Fcol}
\end{equation}
The product of these terms gives the score of a pair, $S_{ij} = P_{ij} \cdot F^{\text{row}}_i \cdot F^{\text{col}}_j$. The scores of the retained pairs are averaged and multiplied by the coverage,
\begin{equation}
S_{\text{total}} = C \cdot \frac{1}{N_{\text{assigned}}} \sum_{(i,j)} S_{ij}. \label{eq:Stotal}
\end{equation}
 
The normalised MI matrices of all 18 models are shown in Fig.~\ref{fig:mi_full}. In each matrix, the entry in a row and a column is the normalised MI between the parameter of that row and the latent variable of that column. A large MI in an outlined cell does not by itself imply a clean mapping, since the same parameter may also be spread over other latent variables (the rest of its row), or the same latent variable may mix several parameters (the rest of its column). These are the two effects penalised by the row and column factors in equations~\eqref{eq:Frow} and~\eqref{eq:Fcol}, so that a clean one-to-one encoding has a large MI in the outlined cell and small values in the rest of its row and column. Hence, the score $S$ above each panel of Fig.~\ref{fig:mi_full} combines both aspects.

We note three patterns in Fig.~\ref{fig:mi_full}. First, only three of the 18 models assign all six parameters to separate latent variables cleanly, that is, with the normalised MI of every matched pair above the threshold: SA lowTTTEEE+full with $L=6$ and 7, and SA lowTTTEEE+high with $L=7$. These are also the three models with the highest scores. Second, in every model the latent variable matched to $\tau$ shares almost no information with $A_s$, and the one matched to $A_s$ shares almost none with $\tau$ (normalised MI of at most 0.03, see Section~\ref{sec:disc_deg}). Third, the SA models have a mean disentanglement score of 0.58 and the CA models of 0.40. The gap is not the same everywhere: for lowTT-TEEE+high the two strategies score almost the same at every $L$, whereas for lowTTTEEE+full, and for lowTTTEEE+high at $L=7$, it is large (Table~\ref{tab:model_comparison}). SA and CA differ only in the loss weight of the disentanglement loss, which is fixed at $w_d=5$ in SA and learned in CA, where it takes values of 0.45--0.81 (Table~\ref{tab:weights}). The gap in the mean disentanglement scores shows the effect of a stronger disentanglement loss weight, which pulls the latent space more towards the physical parameters. These measures indicate how interpretable the latent variables are, but not how well the models fit the real data. Fits to Planck data are hence performed next.

\subsection{Latent Space Posteriors} \label{sec:post}

We sample the posterior distribution of the latent variables $\bm{z}$ of each model with MCMC, using the \texttt{emcee} sampler \citep{Foreman-Mackey:2012any} with 64 walkers. At each step the decoder gives the spectra $(\hat{D}^{TT}_\ell,\hat{D}^{TE}_\ell,\hat{D}^{EE}_\ell)$, which is compared with Planck 2018 data using the \texttt{Plik\_lite} likelihood \citep{Prince:2019hse,Planck:2019nip}. The VAE spectra span $2\le\ell\le2500$ in all three spectra, but the comparison with the data is made over the multipole range that \texttt{Plik\_lite} covers within this span: $2\le\ell\le2500$ for TT, up to the last bin that lies within the VAE range, and $30\le\ell\le1996$ for TE and EE. A sample of the posterior distributions of the latent variables is shown in Appendix~\ref{app:contours}.

\begin{table*}
\centering
\setlength{\tabcolsep}{18pt}
\renewcommand{\arraystretch}{1.2}
\begin{tabular}{llcccccc}
\toprule
\textbf{Configuration} & \textbf{Strategy} & $\mathbf{L}$ & $\mathbf{\hat{\chi}^2}$ & $\mathbf{\Delta}$\textbf{AIC} & $\mathbf{\Delta}$\textbf{BIC} & $\mathbf{\ln B}$ & \textbf{Score} \\
\midrule
\multirow{6}{*}{lowTTTEEE+full} & \multirow{3}{*}{SA} & 5 & 0.9630 & 1.1 & \textbf{0.0} & 0.3 & 0.619 \\
&  & 6 & 0.9607 & 0.7 & 4.0 & \textbf{0.0} & 0.737 \\
&  & 7 & \textbf{0.9578} & \textbf{0.0} & 7.8 & 0.4 & 0.781 \\
\cline{2-8}
& \multirow{3}{*}{CA} & 5 & \textbf{0.9575} & \textbf{0.0} & \textbf{0.0} & \textbf{0.0} & 0.388 \\
&  & 6 & 0.9603 & 2.8 & 7.2 & 3.5 & 0.144 \\
&  & 7 & 0.9621 & 4.9 & 13.7 & 5.5 & 0.399 \\
\midrule
\multirow{6}{*}{lowTTTEEE+high} & \multirow{3}{*}{SA} & 5 & 0.9782 & 11.0 & 6.6 & 10.6 & 0.510 \\
&  & 6 & \textbf{0.9584} & \textbf{0.0} & \textbf{0.0} & \textbf{0.0} & 0.519 \\
&  & 7 & 0.9662 & 5.8 & 10.2 & 5.2 & 0.766 \\
\cline{2-8}
& \multirow{3}{*}{CA} & 5 & 0.9687 & 7.6 & 3.1 & 7.5 & 0.492 \\
&  & 6 & \textbf{0.9546} & \textbf{0.0} & \textbf{0.0} & \textbf{0.0} & 0.433 \\
&  & 7 & 0.9664 & 8.2 & 12.6 & 8.2 & 0.423 \\
\midrule
\multirow{6}{*}{lowTT-TEEE+high} & \multirow{3}{*}{SA} & 5 & 0.9881 & 14.7 & 8.4 & 14.1 & 0.408 \\
&  & 6 & 0.9653 & 1.8 & \textbf{0.0} & 1.0 & 0.451 \\
&  & 7 & \textbf{0.9606} & \textbf{0.0} & 2.6 & \textbf{0.0} & 0.458 \\
\cline{2-8}
& \multirow{3}{*}{CA} & 5 & 0.9713 & 7.0 & 2.6 & 6.3 & 0.384 \\
&  & 6 & \textbf{0.9581} & \textbf{0.0} & \textbf{0.0} & \textbf{0.0} & 0.454 \\
&  & 7 & 0.9614 & 3.0 & 7.4 & 3.2 & 0.471 \\
\bottomrule
\end{tabular}
\caption{Goodness of fit and model comparison for all 18 models, using the MCMC chains and the \texttt{Plik\_lite} likelihood. $\Delta$AIC, $\Delta$BIC and $\ln B$ are measured from the best $L$ within each configuration and strategy, and the lowest values are highlighted in bold. \textit{Score} is the disentanglement score $S_{\rm total}$ of equation~\eqref{eq:Stotal}, computed from the normalised MI matrices of Fig.~\ref{fig:mi_full}.}
\label{tab:model_comparison}
\end{table*}

\begin{table}
\centering
\renewcommand{\arraystretch}{1.2}
\begin{tabular}{lrrr}
\toprule
\textbf{Configuration} & \textbf{$L=5$} & \textbf{$L=6$} & \textbf{$L=7$} \\
\midrule
lowTTTEEE+full & $+4.3$ & $+0.6$ & $-1.0$ \\
lowTTTEEE+high & $+5.6$ & $+2.5$ & $-0.5$ \\
lowTT-TEEE+high & $+10.2$ & $+3.4$ & $-0.8$ \\
\bottomrule
\end{tabular}
\caption{Evidence comparison of the two disentanglement strategies at fixed latent dimension: $\ln B_{\rm CA,SA}=\ln Z_{\rm CA}-\ln Z_{\rm SA}$ for the same configuration. Positive values favour CA and negative values favour SA.}
\label{tab:BF_M_A}
\end{table}

To compare the models fairly, we use the goodness of fit, two model selection criteria and Bayesian evidence, which account for the number of latent variables in different ways. The first measure is the reduced $\chi^2$, $\hat\chi^2=\chi^2_{\min}/(n-L)$, where $\chi^2_{\min}$ is the minimum $\chi^2$ found in the chains, $n=614$ is the number of binned data points used in the \texttt{Plik\_lite} likelihood \citep{Prince:2019hse}, and all $L$ latent variables are counted as free parameters. The second and third are the model selection criteria $\mathrm{AIC}=\chi^2_{\min}+2L$ and $\mathrm{BIC}=\chi^2_{\min}+L\ln n$, which add a fixed penalty for each latent variable.

The Bayesian evidence ($Z$) is estimated from the MCMC samples using the \texttt{MCEvidence} $k$-nearest-neighbour estimator \citep{Heavens:2017afc}. We report two complementary Bayes factors. First, for a fixed strategy and configuration, $\ln B=\ln Z_{\max}-\ln Z$ measures how much a model is disfavoured relative to the model with the highest evidence among $L=5,6,7$. Second, $\ln B_{\rm CA,SA}=\ln Z_{\rm CA}-\ln Z_{\rm SA}$ compares CA with SA at the same $L$ and configuration. The strength of the preference is interpreted with the Jeffreys scale \citep{Jeffreys:1939xee}, in which $\ln B$ between 2.5 and 5 is moderate and above 5 is strong. For the evidence computation we use the same prior for all models: uniform on $[-10,10]$ for every latent variable. No latent variable is given any prior information by any cosmological parameter, so the fit to the Planck data is carried out entirely in the free latent space.

Table~\ref{tab:model_comparison} compares $L=5$, 6 and 7 for each configuration and strategy. No single $L$ is preferred in all six cases, but there is a clear trend. Going from $L=5$ to $L=6$ improves the fit substantially in four of the six cases, and $L=5$ is strongly disfavoured by the evidence ($\ln B>5$) for lowTTTEEE+high and lowTT-TEEE+high. In the two lowTTTEEE+full cases the sixth latent variable does not improve the fit: for SA, $L=5$ and $L=6$ are indistinguishable ($\ln B=0.3$ and 0.0), and for CA, $L=5$ is the best, with $L=6$ moderately disfavoured ($\ln B=3.5$). Going from $L=6$ to $L=7$ gives no significant gain in any case: the BIC never prefers $L=7$, with $L=7$ having the highest evidence only for SA lowTT-TEEE+high, where its advantage over $L=6$ is $\ln B=1.0$, which is inconclusive. Overall, $L=6$ is the best, or within $\ln B\simeq1$ of the best, in five of the six cases. BIC and the evidence select the same $L$ in four cases (CA lowTTTEEE+full, SA lowTTTEEE+high, CA lowTTTEEE+high and CA lowTT-TEEE+high), including all three CA cases. For SA they differ for lowTTTEEE+full (BIC selects $L=5$ and the evidence $L=6$) and for lowTT-TEEE+high (BIC selects $L=6$ and the evidence $L=7$). 

At fixed $L$ (Table~\ref{tab:BF_M_A}), the evidence favours CA over SA at $L=5$ ($\ln B_{\rm CA,SA}=4.3$--$10.2$) and at $L=6$ ($0.6$--$3.4$, from inconclusive to moderate), and favours SA over CA at $L=7$ ($-1.0$ to $-0.5$, weak or inconclusive). Since the $L=7$ models are themselves disfavoured relative to $L=6$ in most cases, we regard the comparison between $L=5$ and $L=6$ as more relevant, in which CA is favoured.

\vspace{-0.5cm}

\section{Discussion} \label{sec:discussion}

The main aim of this work is to find out how many independent degrees of freedom the CMB power spectra (TT, TE and EE) contain, and what they correspond to, without building a cosmological model \textit{a priori} into the network. Standard analyses sample directly in a predefined set of parameters, whereas here the latent variables are free and are compared with the $\Lambda$CDM parameters only after training is completed. We discuss three aspects of the results: the number of latent variables the spectra need, the separation of $\tau$ from $A_s$, and the trade-off between an interpretable latent space and the quality of the fit to real data.

\subsection{The Intrinsic Dimensionality of the CMB}

We train models with three latent dimensions, $L=\{5,6,7\}$. This range is chosen around the six parameters of $\Lambda$CDM, since searching in this neighbourhood could help locate the number of independent degrees of freedom even in this model-independent approach. In principle, the same exercise can be repeated over any range of $L$, and the minimum number of latent variables can be found through statistical measures such as those used here.

A comparison of the fits for $L=5$, 6 and 7 shows how many latent variables the spectra need. In four of the six cases (SA and CA lowTTTEEE+high, and SA and CA lowTT-TEEE+high), $L=5$ is clearly disfavoured by the evidence and $L=7$ gives no significant gain over $L=6$. In these cases, six latent variables are needed to fit the data, and seven give no further improvement. This is the number of free parameters of $\Lambda$CDM, and one more than the five found for data at $\ell\ge30$ \citep{Piras:2025eip,Sun:2025ypd}. In the two lowTTTEEE+full cases the result is different: for SA, $L=5$ and $L=6$ are indistinguishable, and for CA, $L=5$ is the best. A possible explanation for this preference of $L=5$ in this case might be the use of lowTTTEEE twice in the loss function (once in the full spectrum term, once separately), which gives extra weight to the low-$\ell$ data. This gives a hint of how crucial a proper choice of the loss function could be in order to arrive at scientifically valid results. However, this aspect requires further investigations with a series of possible loss functions (both good and bad) and hence we refrain from making any strong comments on this at this point. The CA lowTTTEEE+full model with $L=6$ is also the least interpretable of all the models: only $\tau$ passes the MI threshold, and its score is 0.14 (Fig.~\ref{fig:mi_full}). Once $L=6$ is reached, the seventh latent variable gives no significant improvement, and the evidence decreases for CA lowTTTEEE+full, SA and CA lowTTTEEE+high, and CA lowTT-TEEE+high. For SA lowTTTEEE+full, $L=7$ lowers $\hat{\chi}^2$ slightly and its disentanglement score is higher, while its evidence is not higher ($\ln B=0.4$ against 0.0 at $L=6$). Only for SA lowTT-TEEE+high is the evidence higher at $L=7$, but the difference is inconclusive ($\ln B=1.0$).

The networks are trained on the full spectra, including the low-$\ell$ polarisation, and the Planck comparison tests how well the resulting latent spaces reproduce the observed spectra. The \texttt{Plik\_lite} likelihood includes the low-$\ell$ temperature bins, whose amplitude is not attenuated by reionisation and can be compared with the high-$\ell$ amplitude. So the data carries some information on the split between $A_s$ and $\tau$. A natural interpretation of the step from $L=5$ to $L=6$ is that the low-$\ell$ polarisation in the training spectra provides information on $\tau$ that is independent of $A_s$, so these spectra contain one more independent degree of freedom than at $\ell\ge30$ alone. This agrees with the MI results of Section~\ref{sec:disc_deg}. Since the number of degrees of freedom is not imposed anywhere in the network, the finding that six are needed to fit these spectra is a direct result from the analysis. Whether, and to what extent, the six latent variables correspond to the six $\Lambda$CDM parameters has to be judged from the MI matrices (Section~\ref{sec:disc_phys}).

With five latent variables, one degree of freedom must be merged or dropped, which degrades the fit (Fig.~\ref{fig:recon_resid}). This costs fit for lowTTTEEE+high and lowTT-TEEE+high, where $L=5$ is strongly disfavoured by the evidence, but not for lowTTTEEE+full, where $L=5$ is competitive with $L=6$. Going to $L=7$ brings no meaningful gain: the BIC never prefers $L=7$, and the evidence shows no significant improvement. The four measures ($\hat{\chi}^2$, AIC, BIC and the Bayesian evidence) therefore point in the same direction: they favour $L=6$ over $L=5$ in four of the six cases and find no significant gain beyond $L=6$. They select the same $L$ in all three CA configurations, whereas for SA the fits at neighbouring $L$ are so close that the preferred $L$ depends on the measure considered (Section~\ref{sec:disc_sa_ca}). The Bayesian evidence penalises dimensions that the data do not require, but it depends on the prior of the latent variables, and the latent scale is not the same in different networks, so we do not read differences of order one in $\ln B$ as significant. Overall, $L=6$ is the most consistently favoured value, being the best or within $\ln B\simeq1$ of the best in five of the six cases.

\subsection{Breaking of the \texorpdfstring{$\bm{A_{\mathrm{s}}}$--$\bm{\tau}$}{As - tau} Degeneracy} \label{sec:disc_deg}

A key result of this work is that the latent variables for $\tau$ and $A_s$ share almost no information in any of the 18 models: the MI between them is at most 0.03. In standard model-dependent analyses, $A_s$ and $\tau$ are degenerate in the high-$\ell$ TT spectrum, which depends on them only through the combination $A_s e^{-2\tau}$ (Section~\ref{subsec:TTtheory}). This degeneracy is broken by adding polarisation data \citep{Planck:2018vyg,Zaldarriaga:1996ke}, since $\tau$ is separately constrained by the reionisation bump in the EE spectrum (Section~\ref{subsec:pol}). Our models include $\ell\ge2$ and polarisation, in both the training spectra and the reconstruction loss, and the separation found in their latent space is consistent with the one that including polarisation data provides in standard analyses. It is notable that the separation of these two physical parameters appears as a result of a deep-learning network with neither a cosmological model built into it, nor any of the latent variables assigned to any parameter in advance.

\subsection{What the Latent Variables Capture} \label{sec:disc_phys}

The strength of the mapping differs strongly between parameters (Fig.~\ref{fig:mi_full}). In most models $\omega_{\rm b}$ is matched to the latent variable with the largest MI, and $\tau$, $\omega_{\rm cdm}$ and $h$ typically follow in that order, whereas $n_s$ and, to a lesser extent, $A_s$ most often fall at or below the threshold of 0.10 used for the disentanglement score. This pattern follows how distinctive the imprint of a parameter on the spectra is.

The shape of $C_\ell^{TT}$ changes with angular scale, because different physical processes dominate at different scales \citep{Hu:2001bc,Zaldarriaga:1997ch}. At the largest scales ($\ell \lesssim 30$), the spectrum is nearly flat, i.e. the Sachs--Wolfe plateau \citep{Sachs:1967er}. It is set by the primordial gravitational potential wells, whose amplitude is proportional to $A_s$, together with the late-time decay of these potentials driven by dark energy. Cosmic variance limits how well parameters can be measured at these large scales \citep{Planck:2018vyg}. At intermediate scales ($30 \lesssim \ell \lesssim 1000$), the spectrum shows acoustic oscillations of the photon-baryon fluid. Here, the positions of the peaks give the angular size of the sound horizon $\theta_s$, which is set by the Hubble constant $h$ together with the matter densities. The relative height of the odd and even peaks is sensitive to $\omega_{\rm b}$, and the peak heights are sensitive to $\omega_{\rm cdm}$. At the smallest scales ($\ell \gtrsim 1000$), photon diffusion during recombination washes out the fluctuations, which gives the exponential Silk damping tail \citep{Silk:1967kq}. The optical depth has its own signature in the form of the reionisation bump in the low-$\ell$ EE and TE spectra (Section~\ref{subsec:pol}), and it also lowers the high-$\ell$ amplitude by $e^{-2\tau}$. The baryon and cold dark matter densities and $h$ therefore produce large changes in the spectra in the form of acoustic features, which a latent variable can pick up easily. In contrast, $A_s$ and $n_s$ change the spectra smoothly, as an overall amplitude and a tilt without acoustic features. A latent space that encodes the shape of the spectra therefore has little reason to tie itself to either of them prominently.

This ordering agrees with the temperature-only analysis of \citet{Piras:2025eip}, where $\omega_{\rm b}$ also has the cleanest latent variable, which they identify with the even--odd modulation of the acoustic peaks. Here, only the ordering (instead of specific MI values) is compared since the MI normalisations differ (Section~\ref{sec:mi}). Moreover, in \citet{Piras:2025eip} $\tau$ is only weakly mapped, since in TT alone it enters mainly through $A_se^{-2\tau}$. In our models it is among the best captured parameters, which is expected, since the large-scale polarisation gives it a signature of its own. Their $n_s$ is mapped more clearly in a latent variable mixed with $\omega_{\rm cdm}$, whereas in our CA models, $n_s$ is the weakest captured parameter. In some of our SA models, however, it is captured about as well as $\omega_{\rm cdm}$ and $h$.

At $\ell\ge30$ the spectra depend on $A_s$ and $\tau$ only through $A_se^{-2\tau}$, and five latent variables suffice, for TT alone \citep{Piras:2025eip}, and for binned TT, TE and EE together \citep{Sun:2025ypd}. Adding polarisation at high $\ell$ is hence not seen to prefer an additional independent degree of freedom. The additional degree of freedom in our models (which includes spectra down to $\ell=2$) is associated with the large-scale polarisation, which constrains $\tau$ separately from the amplitude (Section~\ref{sec:disc_deg}).

The largest normalised MI cross-term involves $A_s$ and $n_s$. In several models, in particular in the CA models and in SA lowTT-TEEE+high, the largest entry in the $A_s$ row outside the matched cell lies in the $n_s$ latent variable, and the largest entry in the $n_s$ row lies in the $A_s$ latent variable. In SA lowTTTEEE+full this cross-term is almost absent. Where it appears the entries are small, at most $0.08$, below the threshold of 0.10. Both parameters describe the primordial spectrum ($A_s$ its amplitude and $n_s$ its tilt) and over a limited range of $\ell$ a change of tilt can be partly traded against a change of amplitude. A latent variable may therefore encode a combination of the two rather than each one separately. A weaker version of the same tendency is seen for $\omega_{\rm cdm}$ and $h$, which lie along the direction of the geometric degeneracy between the matter density and $h$ at a fixed angular scale of the acoustic peaks. In the temperature-only analysis of \citet{Piras:2025eip}, the latent variable that follows the peak positions has higher MI with the angular size of the sound horizon $\theta_s$ than with $h$. Since $\theta_s$ is set by $h$ and the matter densities together, a latent variable that follows $\theta_s$ would show only a partial mapping to $h$. The cross-terms hence point to two pairs of parameters whose effects on the spectra partly overlap, $A_s$ with $n_s$ and, more weakly, $\omega_{\rm cdm}$ with $h$.

\subsection{Semi-Automated \texorpdfstring{\textit{vs}}{vs} Completely Automated Disentanglement} \label{sec:disc_sa_ca}

SA and CA differ only in the loss weight of the disentanglement loss, which is fixed at $w_{d}=5.0$ in SA and learned in CA. The learned values lie between 0.45 and 0.81 (Table~\ref{tab:weights}), which are $\sim5$ to $\sim10$ times smaller than the fixed value. The variation of $w_d$ with $L$, and with the configuration, is small compared to this difference. The learned weight is therefore always well below the value fixed by hand, so CA applies a much weaker disentanglement loss. We compare the two strategies in three respects.

\begin{itemize}
\item \emph{Fit to the data.} The evidence favours CA over SA at $L=5$ and $L=6$ and SA over CA at $L=7$ (Table~\ref{tab:BF_M_A}), so the advantage of CA decreases with $L$ and is not present at $L=7$. When each strategy is taken at its own best $L$, $\ln B_{\rm CA,SA}\simeq4.0$, $2.5$ and $2.4$ for lowTTTEEE+full, lowTTTEEE+high and lowTT-TEEE+high respectively. So, CA strategy is still preferred. Because the evidence is compared between different networks with the same latent prior, we regard these differences as indicative.
\item \emph{Interpretability.} In SA strategy the disentanglement score increases with $L$. The difference in the mean score is about $0.09$, $0.23$ and $0.24$ at $L=5$, $6$ and $7$. A possible reason is that a large fixed loss weight costs more fit when there are few latent variables, whereas with more of them the same constraint can be met at little cost. The gain in disentanglement is not tied to a better fit. For SA lowTTTEEE+high, the best fit is at $L=6$ (score $0.52$), but the highest score is at $L=7$ ($0.77$). A good fit therefore neither requires nor excludes a clean mapping of the latent variables onto the cosmological parameters. Moreover, choosing $L$ by the fit alone would not select the most interpretable latent space. With the disentanglement loss, a one-to-one assignment of all six parameters is reached in three of our SA models, all with $L\ge6$. This is different from the more distributed encoding reported for a conditional VAE by \citet{Sun:2025ypd}. However, the measures differ (kernel alignment there, normalised MI here), so the comparison here is qualitative.
\item \emph{Choice of $L$.} For CA, the four measures ($\hat{\chi}^2$, AIC, BIC, B) agree on $L$ in all three configurations, whereas for SA they differ in two of the three. For SA the fits at neighbouring $L$ are close, so the preferred $L$ depends on the measure.
\end{itemize}

SA and CA are therefore two points on a trade-off between interpretability and fit to the data. SA gives clearer disentanglement, especially at larger $L$. CA needs no choice of $w_d$, has the higher evidence at $L=5$ and 6, and gives a consistent preference for $L$. Since CA prefers a weaker disentanglement loss, its latent variables are spread more widely over the parameters, although they fit the Planck data better. This suggests that a latent space less constrained to follow the standard cosmological parameters can fit the Planck data at least as well, and in our comparison better, than one pulled towards them. A possible reading is that the standard parameters, which are chosen for ease of physical interpretation, may not be the combinations that the CMB data constrain most directly.

\section{Conclusions and Outlook} \label{sec:conclusion}

We asked how many latent variables are required to fit the CMB temperature and polarisation spectra when a neural network is given no cosmological model to start from, and to what extent they map onto the six $\Lambda$CDM parameters. Using our \texttt{OCCAMS} (Optimal Components of CMB with Autoencoders and Model Selection) code, a total of 18 VAEs were trained on \texttt{CLASS} spectra of TT, TE and EE for $2\le\ell\le2500$, for three configurations of the reconstruction loss, two ways of treating the weight of the disentanglement loss (semi-automated, SA, and completely automated, CA, strategies), and three different latent dimensions $L=5,6,7$, thereby making our investigation comprehensive. The three configurations were chosen to separate the roles of the temperature and polarisation data and of the low- and high-$\ell$ ranges. The loss weights of the reconstruction and KL divergence terms, and in CA also of the disentanglement term, were set by multi-task uncertainty weighting within a single training run, without a grid search. The latent dimensions have been considered with an expectation that a well-designed VAE model would lead to the number of minimum dimensions close to the physical ones. However, the process can be repeated with any arbitrary number as a viable test case, in order to arrive at the minimum number of dimensions. All models fit the Planck data well, with $\hat\chi^2$ between $0.955$ and $0.988$. The models were compared using $\hat\chi^2$, AIC, BIC and Bayesian evidence, and the latent variables were interpreted through their normalised mutual information (MI) with the six $\Lambda$CDM parameters. Our main results are the following.
\begin{enumerate}
\item The latent variables capture best the parameters that leave distinctive features in the spectra: $\omega_{\rm b}$, then $\tau$, $\omega_{\rm cdm}$ and $h$. The parameters $n_s$ and $A_s$ are captured least by the pipeline, and the small normalised MI cross-terms (smeared) between them suggest that their effects partly overlap.
\item The latent variables for $\tau$ and $A_s$ share almost no information in any of the 18 models (normalised MI of at most 0.03), despite no external bias forced on any latent parameter explicitly \textit{a priori}. This is consistent with the large-scale polarisation separating the two parameters, as it does in standard analyses, and as TT alone at $\ell\ge30$ cannot.
\item In four of the six cases (SA and CA, for lowTTTEEE+high and lowTT-TEEE+high) the fit improves from $L=5$ to $L=6$ and not beyond, so six latent variables are needed, which is the same as the number of free parameters of $\Lambda$CDM. For $\ell\ge30$, earlier analyses find five \citep{Piras:2025eip,Sun:2025ypd}, which matches the single combination $A_se^{-2\tau}$ that replaces $A_s$ and $\tau$ there. The two lowTTTEEE+full cases differ: for SA, $L=5$ and $L=6$ are indistinguishable, and for CA the fit is best at $L=5$, with the choice of this particular loss function being a possible reason for that. The data therefore do not single out one latent dimension in every case, but show a trend to conclude that $L=6$ is the best, or within $\ln B\simeq1$ of the best, in five of the six cases, and $L=7$ gives no significant gain over $L=6$ in any case.
\item SA and CA strategies are two points on a trade-off between interpretability and fit to the data. SA gives clearer disentanglement, with a mean score of $0.58$ against $0.40$ for CA, and the difference grows with $L$. All six parameters are matched cleanly to a latent variable only in three SA models, all with $L\ge6$. CA removes the need to fix the weight of the disentanglement loss in advance. It also gives a higher evidence at $L=5$ and $6$, and remains preferred when each strategy is compared at the value of $L$ that suits it best. The four model-comparison measures also agree on the preferred $L$ in all three CA cases, which they do not for SA. The gain in disentanglement is not tied to a better fit: for SA lowTTTEEE+full and lowTTTEEE+high the score is highest at $L=7$, whereas the evidence is highest at $L=6$.
\end{enumerate}

These results show that a deep-learning approach can produce a compact parametrisation of the CMB temperature and polarisation spectra, with the loss weights set automatically. No cosmological model is built into the network or its latent space, although the training spectra are generated within $\Lambda$CDM. Starting from this freedom, six latent variables suffice in five of the six cases studied in this work. This is as many as there are free parameters in $\Lambda$CDM, and one more than required for spectra at $\ell\ge30$. The latent variables map onto the $\Lambda$CDM parameters to different extents: more clearly for $\omega_{\rm b}$, $\tau$, $\omega_{\rm cdm}$ and $h$, and less for $A_s$ and $n_s$.

This framework can be extended in several directions. The first is to CMB data of higher precision and wider multipole range. Ground-based experiments such as ACT \citep{AtacamaCosmologyTelescope:2025blo} and SPT-3G \citep{SPT-3G:2014dbx} already measure the spectra to higher $\ell$ than Planck. The Simons Observatory \citep{SimonsObservatory:2025wwn} will improve the small-scale polarisation, and LiteBIRD \citep{LiteBIRD:2026gdf}, a full-sky polarisation mission designed mainly to search for primordial B-modes, will also measure the large-scale polarisation that constrains $\tau$. Since the separation of $\tau$ and $A_s$ in our latent variables is expected to depend on the large-scale polarisation, such data would test it directly, and would show how the number of latent variables and the parameter mappings respond to better measurements.

The second is to apply the framework to baryon acoustic oscillations (BAO) and other late-time data, for example from the Dark Energy Spectroscopic Instrument (DESI) \citep{DESI:2024mwx,DESI:2025zgx}, and to combine them with the CMB latent variables. This would allow extensions, such as dynamical dark energy, to be studied in a data-driven parametrisation, as was done for early dark energy in the temperature-only case by \citet{Piras:2025eip}. Applying the same pipeline to spectra from extended models would also show whether the preferred latent dimension follows the underlying model. Within the CMB, the latent variables can also be used directly in parameter inference and related to derived combinations of physical parameters such as $\theta_s$.

Finally, the architecture and training strategy used here are not final. The VAE, the multi-task weighting and the automated choice of the disentanglement weight can be refined further. We plan to improve the statistical robustness of these techniques, and hope that this work encourages the development of more sophisticated, reliable and widely applicable data-driven approaches in cosmology.

\section*{Acknowledgments}

We acknowledge the computational facilities provided by the SyMeC High-Performance Computing (HPC) cluster of Indian Statistical Institute (ISI) Kolkata and the Pegasus HPC facility at the Inter-University Centre for Astronomy and Astrophysics (IUCAA), Pune, India. AM and RS thank Tanima Duary, Purba Mukherjee, Biswajit Das, Debarun Paul and Arko Bhaumik, for insightful discussions. AM thanks the Anusandhan National Research Foundation (ANRF), Government of India and ISI Kolkata for financial support through Junior Research Fellowship. RS thanks ISI Kolkata for financial support through Senior Research Fellowship. MWH, SS, AAS and SP thank ANRF, Government of India, for partial support through Project No. CRG/2023/003984. SP also thanks IUCAA, Pune for Visiting Associateship during which a part of the work was done.

\section*{Data Availability}
The \texttt{OCCAMS} (Optimal Components of CMB with Autoencoders and Model Selection) suite and data for this article will be made public on GitHub once the paper is published. All other data sets used in this work are publicly available. The modified versions of the codes by other authors may be made available upon reasonable request.

\bibliographystyle{mnras}
\bibliography{biblio}

\appendix

\section{Construction of \texorpdfstring{$\bm{\mathcal{L}}_{\bm{\mathrm{\MakeLowercase{dis}}}}$}{L-dis}}
\label{app:L_dis}

\begin{figure}
\centering
\begin{tikzpicture}[
>=stealth,
every node/.style={font=\scriptsize},
matrix node/.style={
matrix of math nodes,
left delimiter=[,
right delimiter=],
nodes={minimum width=0.62cm, minimum height=0.46cm, inner sep=1pt, anchor=center}
}
]
\matrix (ZT) [matrix node]
{
\theta_{1,1} & \theta_{1,2} & \cdots & \theta_{1,N} \\
\theta_{2,1} & \theta_{2,2} & \cdots & \theta_{2,N} \\
\vdots  & \vdots  & \ddots & \vdots  \\
\theta_{P,1} & \theta_{P,2} & \cdots & \theta_{P,N} \\
};
\node[above=0.15cm of ZT] {\textbf{Transposed Parameters ($\tilde\Theta^T$)}};
\node[left=0.22cm of ZT-1-1, font=\tiny, text=blue!80!black, align=right] (labeltau) {Param 1 ($\tau$)\\$\rightarrow$};
\node[left=0.22cm of ZT-2-1, font=\tiny, align=right] {Param 2 ($A_s$)\\$\rightarrow$};
\node (times) [right=0.1cm of ZT] {$\times$};
\matrix (Z) [matrix node, right=0.1cm of times]
{
z_{1,1} & z_{2,1} & \cdots & z_{L,1} \\
z_{1,2} & z_{2,2} & \cdots & z_{L,2} \\
\vdots  & \vdots  & \ddots & \vdots  \\
z_{1,N} & z_{2,N} & \cdots & z_{L,N} \\
};
\node[above=0.15cm of Z] {\textbf{Latent Batch ($\tilde Z$)}};
\node[above=0.0cm of Z-1-2, font=\tiny, text=red!80!black] (labelAs) {$\downarrow$ Latent 2};
\matrix (Cov) [matrix node, below=0.9cm of ZT]
{
R_{1,1} & R_{1,2} & \cdots & R_{1,L} \\
R_{2,1} & R_{2,2} & \cdots & R_{2,L} \\
\vdots       & \vdots       & \ddots & \vdots       \\
R_{P,1} & R_{P,2} & \cdots & R_{P,L} \\
};
\node[above=0.15cm of Cov] {\textbf{Squared Correlation Matrix ($R$)}};
\node (equals) [left=0.3cm of Cov, font=\normalsize] {$=$};
\begin{scope}[on background layer]
\node[fill=blue!10, rounded corners, fit=(ZT-1-1)(ZT-1-4), inner sep=0pt] (row1) {};
\node[fill=red!10, rounded corners, fit=(Z-1-2)(Z-4-2), inner sep=0pt] (col2) {};
\node[fill=purple!20, draw=purple, thick, rounded corners, fit=(Cov-1-2), inner sep=1pt] (cell12) {};
\end{scope}
\node (dotprod) [below=0.8cm of Z, xshift=0.25cm, text width=3.7cm, align=center, draw, rounded corners, inner sep=4pt, anchor=north] {
\textbf{1. The Dot Product (Correlation)} \\
\vspace{0.1cm}
$R_{1,2} = \Big(\tfrac{1}{N}\sum_{k=1}^N (\tilde\Theta^T)_{1,k} (\tilde Z)_{k,2}\Big)^2$ \\
\vspace{0.1cm}
\scriptsize{This cell is the squared correlation between parameter 1 ($\tau$) and latent 2 across the batch of $N$ samples. Dividing each column of $R$ by its sum gives $\mathbf M$, whose columns sum to 1.}
};
\node (loss) [below=1.1cm of Cov, xshift=2.1cm, text width=6.6cm, align=center, draw, thick, rounded corners, fill=gray!10, inner sep=4pt, anchor=north] {
\textbf{2. Optimisation (The Penalty)} \\
\vspace{0.1cm}
$\mathcal{L}_{\text{dis}} = \mathcal{L}_{\text{purity}} + \alpha\,\mathcal{L}_{\text{collision}} + \beta'\mathcal{L}_{\text{strict}}$ \\
\vspace{0.1cm}
\scriptsize{Purity and strict act on the columns of $\mathbf M$. The collision term, $\sum_{i\neq j}(\mathbf M^T\mathbf M)_{i,j}^2$, penalises latents that carry the same parameters, so that each latent tracks few parameters and different latents track different ones.}
};
\draw[->, thick, purple, dashed] (cell12.south) |- (dotprod.west);
\draw[->, thick] (dotprod.south) -- (loss.north);
\end{tikzpicture}
\caption{Visualisation of the disentanglement penalty calculation during a training epoch. The network multiplies the transposed standardised parameter batch ($\tilde\Theta^T$) by the standardised latent batch ($\tilde Z$). Each element $R_{i,j}$ is the squared correlation between parameter $i$ (e.g. $\tau$) and latent $j$. Normalising each column of $R$ to sum to one gives $\mathbf M$, from which the three terms of $\mathcal{L}_{\text{dis}}$ are computed: purity and strict from the columns of $\mathbf M$, and collision from the off-diagonal elements of $\mathbf M^T\mathbf M$. Minimising these terms makes each latent follow only a few parameters, and different latents follow different ones.}
\label{fig:matrix_multiplication}
\end{figure}
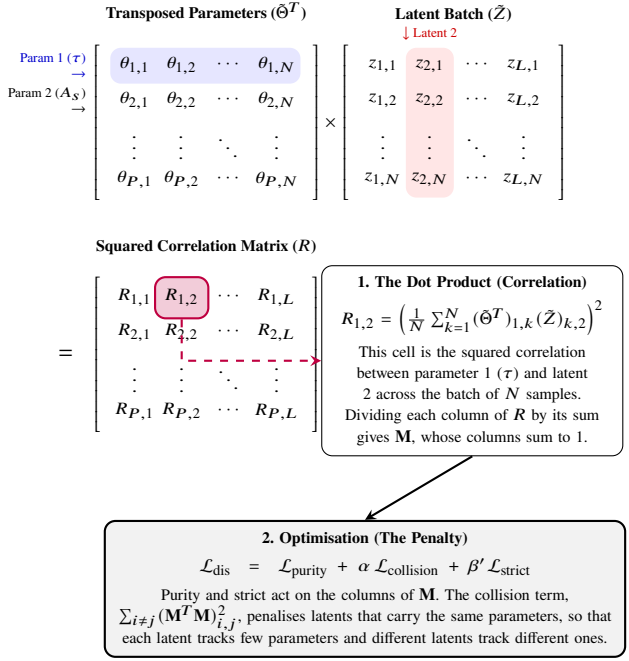

Given a batch of $N$ samples, let $\mathbf{Z} \in \mathbb{R}^{N \times L}$ be the latent representations and $\mathbf{\Theta} \in \mathbb{R}^{N \times P}$ be the theoretical model parameter labels, where $L$ is the latent dimension of a particular VAE model and $P$ is number of cosmological model parameters (6 for $\Lambda$CDM). We first apply zero-mean, unit-variance standardisation to obtain $\tilde{\mathbf{Z}}$ and $\tilde{\mathbf{\Theta}}$. 

Then the squared Pearson correlation matrix is computed between each latent dimension and each physical parameter. To frame this as a probability distribution over the parameters for each latent node, we apply column-wise normalisation to obtain the matrix $\mathbf{M} \in \mathbb{R}^{P \times L}$,
\begin{equation*}
\mathbf{M}_{i,j} = \frac{(\tilde{\mathbf{\Theta}}_{:,i} \cdot \tilde{\mathbf{Z}}_{:,j})^2 / N^2}{\sum_{k=1}^P (\tilde{\mathbf{\Theta}}_{:,k} \cdot \tilde{\mathbf{Z}}_{:,j})^2 / N^2 + \epsilon},
\end{equation*}
where $\epsilon = 10^{-8}$ ensures numerical stability. The columns of $\mathbf{M}$ sum to $1$, representing the fractional information capacity each latent variable dedicates to the respective parameters. Fig.~\ref{fig:matrix_multiplication} summarises the construction of $\mathbf M_{i,j}$ and of the loss terms as follows. To enforce 1-to-1 mappings while remaining sensitive to possible physical degeneracies (e.g., $A_s e^{-2\tau}$), our total disentanglement loss is defined as a weighted sum of three distinct objectives,
\begin{equation*}
\mathcal{L}_{\text{dis}} = \mathcal{L}_{\text{purity}} + \alpha \mathcal{L}_{\text{collision}} + \beta' \mathcal{L}_{\text{strict}}.
\end{equation*}
\textbf{1. Top-2 Purity Loss ($\mathcal{L}_{\text{purity}}$):} We penalise latent variables that distribute information across more than two parameters. By extracting the top two values in each column of $\mathbf{M}$ (denoted as $v_{1,j}$ and $v_{2,j}$ for the $j$-th latent variable), we encourage the network to concentrate its predictive power,
\begin{equation*}
\mathcal{L}_{\text{purity}} = \sum_{j=1}^L \left( 1 - (v_{1,j} + v_{2,j}) \right)^2.
\end{equation*}

\textbf{2. Collision Loss ($\mathcal{L}_{\text{collision}}$):} To prevent mode collapse where multiple latent variables track the same "easy" parameter, we compute the overlap (Gram) matrix $\mathbf{O} = \mathbf{M}^T \mathbf{M} \in \mathbb{R}^{L \times L}$. We penalise the off-diagonal elements of $\mathbf{O}$, encouraging orthogonality between different latent dimensions,
\begin{equation*}
\mathcal{L}_{\text{collision}} = \sum_{i=1}^L \sum_{\substack{j=1 \\ j \neq i}}^L\mathbf{O}_{i,j}^2.
\end{equation*}

\textbf{3. Strict Purity Tie-Breaker ($\mathcal{L}_{\text{strict}}$):} To ensure the network prefers perfect 1-to-1 mappings and only utilises the 2-parameter degeneracy allowance only when physically necessary to lower the reconstruction error, we apply a minor penalty to the deviation of the absolute maximum column value from 1,
\begin{equation*}
\mathcal{L}_{\text{strict}} = \sum_{j=1}^L \left( 1 - \max_{i}(\mathbf{M}_{i,j}) \right)^2.
\end{equation*}

The hyperparameters $\alpha$ and $\beta'$ control the trade-off between orthogonality and strictness. In practice, $\beta' \ll 1.0$ acts as a soft regulariser against unnecessary use of the two-parameter allowance.

\section{Loss Function Weights}\label{app:weights}

Table~\ref{tab:weights} lists the loss weights at the end of training for all 18 models, as defined in equation~\eqref{eq:generic_loss}. The low-$\ell$ weights are larger than the high-$\ell$ weights by roughly two orders of magnitude, and $\beta$ lies between 0.15 and 0.26 in all models.

\begin{table}
\centering
\setlength{\tabcolsep}{3pt}
\renewcommand{\arraystretch}{1.1}
\begin{tabular}{ccccccc}
\toprule
\textbf{Configuration} & \textbf{Strategy} & \textbf{$L$} & \textbf{$\mathbf{w_l\,(10^4)}$} & \textbf{$\mathbf{w_h\,(10^2)}$} & \textbf{$\mathbf{\beta}$} & \textbf{$\mathbf{w_d}$} \\
\midrule
\multirow{6}{*}{lowTTTEEE+full} & \multirow{3}{*}{SA} & 5 & 0.168 & 0.283 & 0.253 & 5 \\
 &  & 6 & 0.994 & 1.16 & 0.209 & 5 \\
 &  & 7 & 0.973 & 1.21 & 0.238 & 5 \\
\cline{2-7}
 & \multirow{3}{*}{CA} & 5 & 0.346 & 0.225 & 0.238 & 0.810 \\
 &  & 6 & 1.28 & 1.21 & 0.191 & 0.587 \\
 &  & 7 & 1.69 & 1.27 & 0.192 & 0.446 \\
\midrule
\multirow{6}{*}{lowTTTEEE+high} & \multirow{3}{*}{SA} & 5 & 0.178 & 0.272 & 0.250 & 5 \\
 &  & 6 & 0.817 & 1.13 & 0.212 & 5 \\
 &  & 7 & 1.61 & 1.24 & 0.184 & 5 \\
\cline{2-7}
 & \multirow{3}{*}{CA} & 5 & 0.323 & 0.22 & 0.239 & 0.799 \\
 &  & 6 & 1.19 & 1.17 & 0.199 & 0.583 \\
 &  & 7 & 1.64 & 1.24 & 0.193 & 0.454 \\
\midrule
\multirow{6}{*}{lowTT-TEEE+high} & \multirow{3}{*}{SA} & 5 & 1.07, 0.274 & 0.159 & 0.190 & 5 \\
 &  & 6 & 1.58, 1.27 & 1.06 & 0.160 & 5 \\
 &  & 7 & 2.56, 2.35 & 1.24 & 0.151 & 5 \\
\cline{2-7}
 & \multirow{3}{*}{CA} & 5 & 1.33, 0.354 & 0.134 & 0.185 & 0.799 \\
 &  & 6 & 1.80, 1.42 & 1.17 & 0.162 & 0.596 \\
 &  & 7 & 1.74, 1.75 & 1.24 & 0.165 & 0.448 \\
\bottomrule
\end{tabular}
\caption{Loss weights at the end of training for all 18 models. $w_l$ and $w_h$ are the weights of the low-$\ell$ and high-$\ell$ reconstruction terms, $\beta$ is that of the KL term, and $w_d$ is that of the disentanglement term. For lowTT-TEEE+high the $w_l$ column lists $(w_{\rm lowTT},w_{\rm lowTEEE})$, and for lowTTTEEE+full the $w_h$ column holds $w_{\rm full}$.}
\label{tab:weights}
\end{table}

\section{Disentanglement Score Dependence on MI Threshold}
\label{app:threshold}

\begin{figure}
\centering
\includegraphics[width=0.48\textwidth]{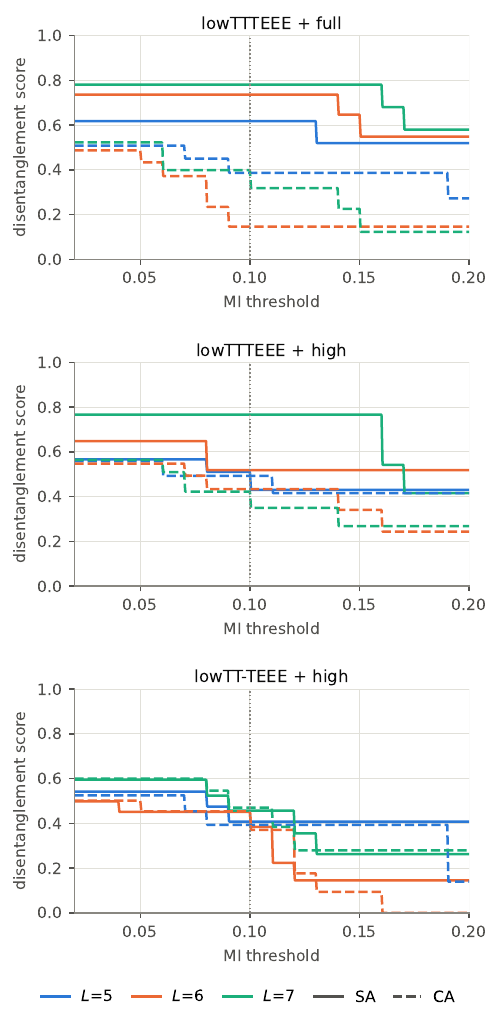}
\caption{Sensitivity of the disentanglement score to the MI threshold for the three configurations (solid: SA, dashed: CA; colours: $L$). The dotted line marks the nominal threshold of 0.10 used in the main analysis.}
\label{fig:mi_threshold}
\end{figure}

The disentanglement score of Section~\ref{sec:results} counts a parameter as captured only if its matched MI exceeds the threshold $M_{\rm min}=0.10$. Fig.~\ref{fig:mi_threshold} shows the score of all 18 models as a function of this threshold between 0.02 and 0.20. Each curve is a step function, because the score drops whenever the threshold passes the matched MI of one parameter and that parameter leaves the coverage. Below about 0.04 all scores sit on their plateau values. The three highest-scoring models at the nominal threshold, SA lowTTTEEE+full with $L=6$ and $7$ and SA lowTTTEEE+high with $L=7$, remain the three highest up to a threshold of 0.17; above it no model exceeds a score of 0.58. The CA models lose a larger part of their score at thresholds between 0.05 and 0.10 than the SA models, so their scores depend more strongly on the chosen threshold. The ordering among the lower-scoring models changes with the threshold, and hence differences between them should not be over-interpreted.

\vspace{-0.5cm}

\section{MCMC Contours} \label{app:contours}

\begin{figure}
\centering
\includegraphics[width=\columnwidth]{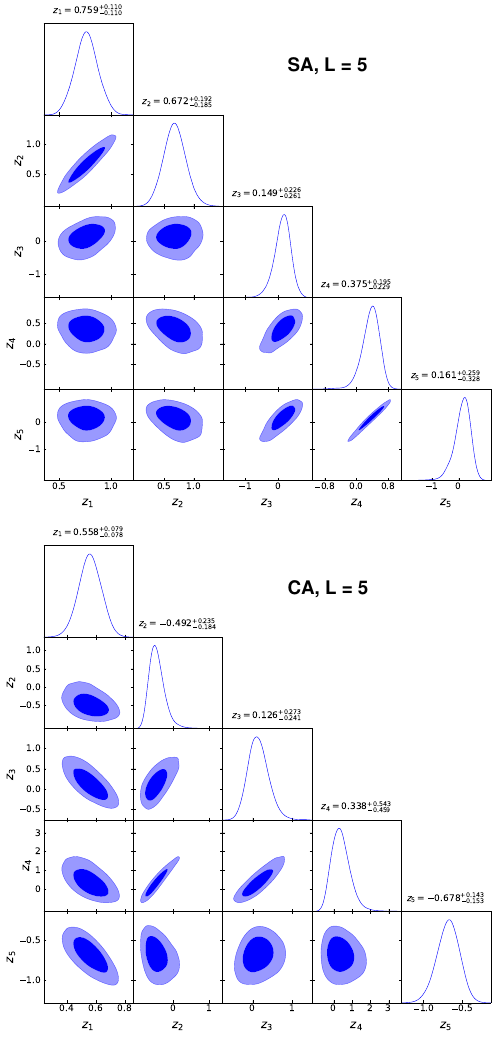}
\caption{Posterior distributions of the latent variables for the \mbox{lowTTTEEE+high} models with $L=5$: SA (top) and CA (bottom). Contours enclose 68 and 95 per cent of the posterior.}
\label{fig:contours_L5}
\end{figure}

\begin{figure}
\centering
\includegraphics[width=\columnwidth]{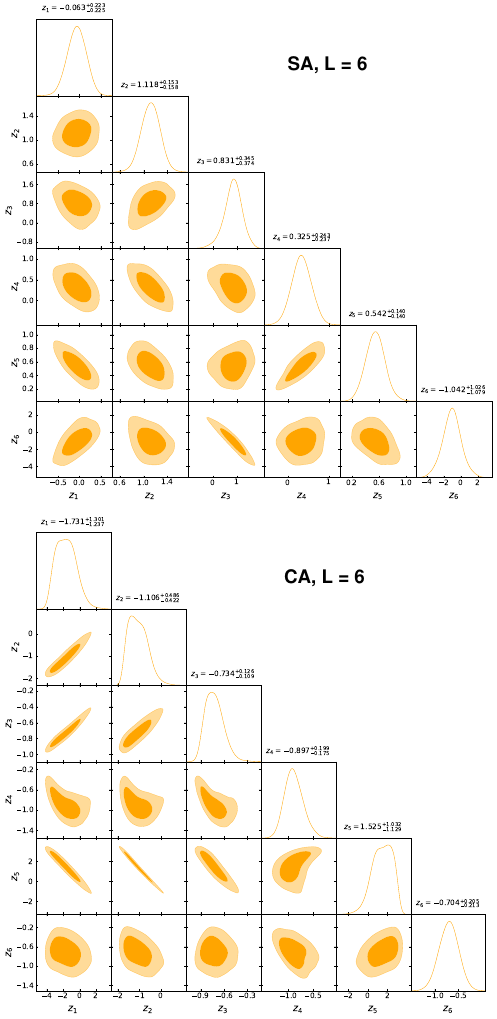}
\caption{Posterior distributions of the latent variables for the \mbox{lowTTTEEE+high} models with $L=6$: SA (top) and CA (bottom). Contours enclose 68 and 95 per cent of the posterior.}
\label{fig:contours_L6}
\end{figure}

\begin{figure}
\centering
\includegraphics[width=\columnwidth]{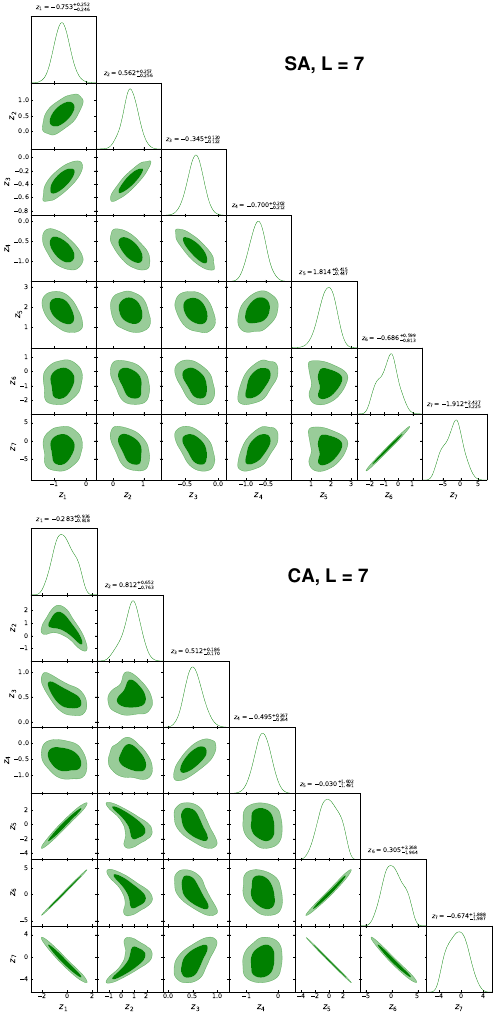}
\caption{Posterior distributions of the latent variables for the \mbox{lowTTTEEE+high} models with $L=7$: SA (top) and CA (bottom). Contours enclose 68 and 95 per cent of the posterior.}
\label{fig:contours_L7}
\end{figure}

Figs.~\ref{fig:contours_L5}, \ref{fig:contours_L6} and \ref{fig:contours_L7} show the marginal posteriors of the latent variables $\bm{z}$ for the \mbox{lowTTTEEE+high} models, plotted using \texttt{getdist} \citep{Lewis:2019xzd}. The numbering, sign and scale of the latent variables are arbitrary, so the panels cannot be compared latent by latent. For $L=6$ and 7, some pairs of latent variables are almost perfectly (anti-)correlated, indicating that the data constrain only a combination of the two and not each separately.

\clearpage
 
\bsp	
\label{lastpage}

\end{document}